\documentclass[journal]{IEEEtran}

\usepackage{cite}
\usepackage{hyperref}

\ifCLASSINFOpdf
    \usepackage[pdftex]{graphicx}
    \graphicspath{{./figures/}}
\else
\fi

\usepackage{mathtools}
\usepackage{algorithm}
\usepackage{algorithmic}
\usepackage{eqparbox}

\usepackage{xcolor}
\usepackage{xspace} 

\ifCLASSOPTIONcompsoc
    \usepackage[caption=false,font=normalsize,labelfont=sf,textfont=sf]{subfig}
\else
  \usepackage[caption=false,font=footnotesize]{subfig}
\fi
\usepackage{siunitx}
\usepackage{mathptmx}
\usepackage{threeparttable}
\usepackage{amsmath,amssymb,amsfonts}
\usepackage{algorithmic}
\usepackage{textcomp}
\usepackage{array}
\usepackage{multirow}

\begin{document}
\bstctlcite{IEEEexample:BSTcontrol} 

\title{A Miniature Batteryless Bioelectronic Implant Using One Magnetoelectric Transducer for Wireless Powering and PWM Backscatter Communication}

\author{
        Zhanghao~Yu\textsuperscript{*},~\IEEEmembership{Member,~IEEE,}
        Yiwei~Zou\textsuperscript{*},~\IEEEmembership{Student Member,~IEEE,}
        Huan-Cheng~Liao,~\IEEEmembership{Student Member,~IEEE,}
        Fatima~Alrashdan,~\IEEEmembership{Member,~IEEE,}
        Ziyuan~Wen,~\IEEEmembership{Student Member,~IEEE,}
        Joshua~E.~Woods,~\IEEEmembership{Student Member,~IEEE,}
        Wei~Wang,~\IEEEmembership{Student Member,~IEEE,}
        Jacob~T.~Robinson,~\IEEEmembership{Senior~Member,~IEEE,}
        and~Kaiyuan~Yang,~\IEEEmembership{Member,~IEEE}

\thanks{Manuscript received on}
\thanks{*Z. Yu and Y. Zou contributed equally to this work.}
\thanks{All authors are with the Department of Electrical and Computer Engineering, Rice University, Houston, TX 77005, USA. 
Z. Yu is now with Intel, Santa Clara, CA 95054, USA. 
(Corresponding author: Kaiyuan Yang, kyang@rice.edu)}
\thanks{This work was supported in part by the National Science Foundation (NSF) awards 2023849 and 2146476, the Robert and Janice McNair Foundation, McNair Medical Institute, the National Institutes of Health (NIH) under grant U18EB029353, and the Defense Advanced Research Projects Agency (DARPA) under Agreement FA8650-21-2-7119. F. Alrashdan, J. E. Woods, J. T. Robinson, and K. Yang receive monetary and/or equity compensation from Motif Neurotech.}
}



\maketitle

\begin{abstract}
Wireless minimally invasive bioelectronic implants enable a wide range of applications in healthcare, medicine, and scientific research. 
Magnetoelectric (ME) wireless power transfer (WPT) has emerged as a promising approach for powering miniature bio-implants because of its remarkable efficiency, safety limit, and misalignment tolerance. However, achieving low-power and high-quality uplink communication using ME remains a challenge. This paper presents a pulse-width modulated (PWM) ME backscatter uplink communication enabled by a switched-capacitor energy extraction (SCEE) technique. The SCEE rapidly extracts and dissipates the kinetic energy within the ME transducer during its ringdown period, enabling time-domain PWM in ME backscatter. Various circuit techniques are presented to realize SCEE with low power consumption. This paper also describes the high-order modeling of ME transducers to facilitate the design and analysis, which shows good matching with measurement. 
Our prototyping system includes a millimeter-scale ME implant with a fully integrated system-on-chip (SoC) and a portable transceiver for power transfer and bidirectional communication. 
SCEE is proven to induce \textgreater 50\% amplitude reduction within 2 ME cycles, leading to a PWM ME backscatter uplink with 17.73 kbps data rate and 0.9 pJ/bit efficiency. It also achieves 8.5$\times$10\textsuperscript{-5} bit-error-rate (BER) at a 5 cm distance, using a lightweight multi-layer-perception (MLP) decoding algorithm. Finally, the system demonstrates continuous wireless neural local-field potential (LFP) recording in an \textit{in vitro} setup. 
\end{abstract}


\section{Introduction}

\IEEEPARstart 
{I}{mplantable} bioelectronics present a paradigm shift in clinical interventions and monitoring, heralding a new era in medicine and patient care. Wireless power transfer technologies eliminate the need for bulky batteries and wires in bioelectronic implants. This miniaturization, down to the (sub-)millimeter scale, facilitates minimally invasive implantation, reduces the risk of infection, and enhances targeting specificity \cite{nair_miniature_2023,gao_wearable_2021,mickle_wireless_2019}. Additionally, compared to today's battery-powered implants, wirelessly powered devices offer extended operational lifespans.

\begin{figure}[t!]
\centering
\includegraphics[width = \linewidth]{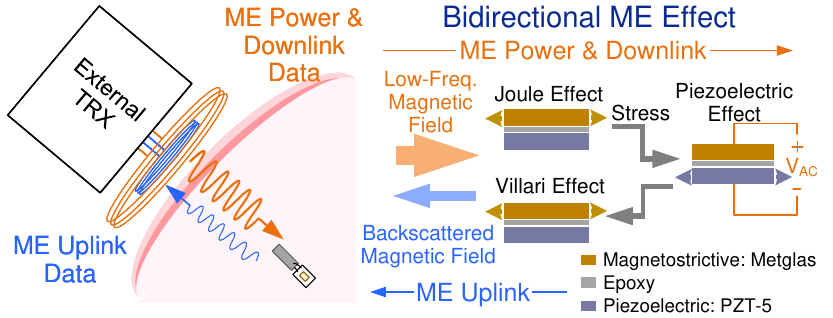}

\caption{Conceptual diagram of a magnetoelectric (ME) implant using a single ME transducer for power and bidirectional data transfers.}
\label{ME_WPT}
\end{figure}

Despite the immense potential of wireless battery-less implants, two critical engineering hurdles must be overcome: ensuring safe and reliable wireless power transfer (WPT) and establishing efficient bidirectional telemetry. Researchers have studied and demonstrated various WPT modalities, including inductive coupling~\cite{jia_mm-sized_2019, lee_neural_2021, kwon_battery-less_2023}, ultrasound~\cite{seo_wireless_2016, jiang_flexible_2022, ghanbari_sub-mm3_2019,piech_wireless_2020}, light~\cite{lim_light-tolerant_2022, lee_sub-mm3_2024,lee_330m90m_2018}. However, each modality presents trade-offs: balancing the implant receiver's size, the maximum power delivery under safety limits, and the system's tolerance to misalignment between the implant and the external power transmitter.
%
An emerging WPT technology using magnetoelectric (ME) effects stands out due to its superior power transfer efficiency to a millimeter-scale receiver, high power delivery within safety limits, and lower sensitivity to misalignment \cite{yu_magni_2020, chen_wireless_2022, yu_wireless_2022, hosur_magsonic_2023, woods_miniature_2024, wang_171_2024}. The ME effect converts low-frequency magnetic fields to electrical energy via acoustic coupling \cite{zhai_giant_2006}, as shown in Fig.~\ref{ME_WPT}.
This low operating frequency results in reduced body absorption and reflection, allowing deeper tissue penetration compared to other WPT methods. A previous work with ME power transfer \cite{yu_magnetoelectric_2022} shows that an operating depth up to 60 mm (0.1 mT magnetic field) can be achieved without violating the electrical field limit and the specific absorption rate (SAR) limit in IEEE standard (unrestricted) \cite{noauthor_ieee_2019}. 
Furthermore, attributed to the high permeability of the magnetostrictive materials, the magnetic flux concentration effect greatly enhances robustness to angular misalignment \cite{yu_magnetoelectric_2022}. These features make ME a compelling WPT solution for millimeter-scale implantable bioelectronics.




\begin{figure}[t!]
\centering
\includegraphics[width = 0.95\linewidth]{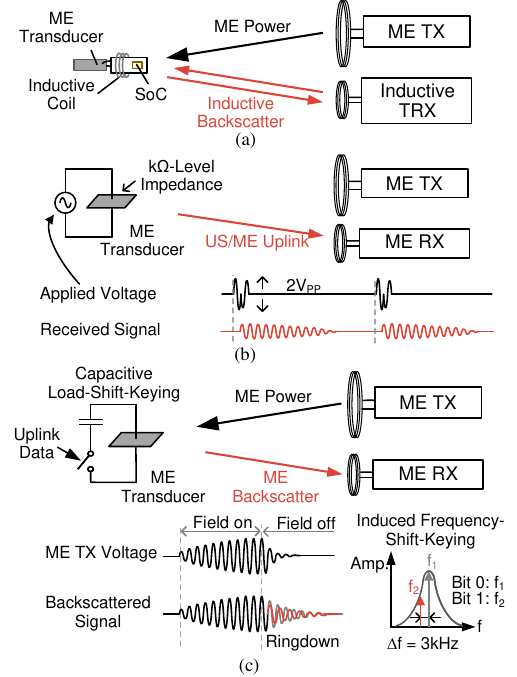}
\caption{Existing uplink technologies for ME implants, including (a) the hybrid scheme~\cite{charthad_mm-sized_2015}; (b) active driving of an ME transducer~\cite{hosur_short-range_2023, hosur_magsonic_2023}; (c) ME backscatter with load-shift-keying~\cite{yu_wireless_2022}.}
\label{Scheme_comparison}
\end{figure}

Meanwhile, bidirectional wireless telemetry is essential for real-time sensing and closed-loop physiological control, which must be co-designed with the WPT scheme to achieve optimal performance and device miniaturization. 
Integrating a second antenna, most commonly an inductive coil~\cite{charthad_mm-sized_2015,yu_wireless_2022}, is the most straightforward uplink solution with a high data rate, as shown in Fig.~\ref{Scheme_hybrid}. However, this approach increases the implant size, complicates device assembly, and introduces interference between power and telemetry channels. Instead, sharing a single transducer for power and data is highly desirable for device miniaturization and system efficiency. For common WPT modalities, such as inductive coupling~\cite{zhao_wireless_2024} and ultrasound~\cite{ghanbari_sub-mm3_2019,piech_wireless_2020}, low-power backscatter communication technique can be readily applied by modulating the load impedance to the transducer. Uplink communication with ME transducers is more challenging due to its non-reciprocal properties. Recent studies \cite{hosur_short-range_2023,hosur_magsonic_2023, yu_magnetoelectric_2022} revealed possible solutions based on converse ME effects, which refers to the phenomenon where vibrations in the magnetostrictive layer lead to changes in the material magnetization. In \cite{hosur_short-range_2023}, uplink telemetry is realized by actively driving an ME transducer. Specifically, the transducer is driven with two cycles of sine waves with opposite phases, and pulse position encoding is used for data modulation, as shown in Fig.~\ref{Scheme_driving}. In \cite{hosur_magsonic_2023}, the active driving scheme is extended to high-frequency resonance modes to demonstrate simultaneous power transfer and uplink. While active driving achieves a high data rate, driving the k$\Omega$-impedance of ME transducers with large amplitude consumes significant power in order to generate a strong enough received signal.  In a separate study~\cite{yu_magnetoelectric_2022}, researchers discovered that modulating the capacitive or resistive loading of an ME transducer during its ringdown period modulates its vibration frequency without burning power, effectively achieving ultra-low-power backscatter communication. Based on this principle, an ME backscatter telemetry using load shift keying (LSK) is demonstrated (see Fig.~\ref{Scheme_lsk}). In experiments, the high-Q (quality factor) ME transducer continues to vibrate for more than 30 cycles after turning off the excitation magnetic field, during which the capacitive loading is switched according to uplink data. Despite the advancement, LSK ME backscatter encounters limited SNR and data rate. Due to the high-Q nature of the ME transducer, a direct trade-off between frequency shift and signal strength caps the SNR achievable at the receiver end. Besides, the excitation and ringdown processes take tens of ME cycles to transmit a single bit, limiting the achievable data rate. 
%
%

\begin{figure}[t!]
\centering
\includegraphics[width = \linewidth]{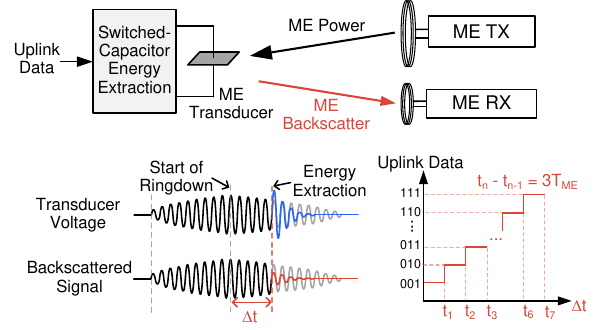}

\caption{Principle and waveforms of the proposed PWM ME backscatter.}
\label{Scheme_PWM}
\end{figure}

To improve the reliability and data rate of uplink telemetry with an ME transducer, this paper presents a pulse-width modulated (PWM) ME backscatter scheme enabled by a switched-capacitor energy extraction (SCEE) technique~\cite{yu_millimetric_2024}. SCEE quickly dissipates energy in an ME transducer and induces an abrupt amplitude reduction in the backscattered signal for PWM, as illustrated in Fig.~\ref{Scheme_PWM}. 
Compared to FSK modulation in \cite{yu_magnetoelectric_2022}, time-domain PWM backscatter improves both SNR and data rate because it avoids deviating from the ME transducer's resonant frequency and transmits multiple bits in a single ringdown period, amortizing the excitation and ringdown time.
To better analyze and optimize SCEE-enabled ME backscatter, we further report a high-order electrical model for ME transducers. This model accounts for the high-frequency resonance modes of the ME transducer, attaining excellent matching between simulation and measurement.
We developed and evaluated a prototyping system including a 6.7 mm$^3$ ME implant and a custom portable transceiver (TRX) for wireless recording and stimulation. The implant SoC integrates power management, bidirectional data transmission, bio-stimulation, temperature sensing, and LFP/ECoG/EGM/ECG recording, targeting applications in cortical and cardiac implants. Overall, the system demonstrates (1) wireless power, downlink, and PWM backscattering uplink using a single 5$\times$2$\times$0.2 mm$^3$ ME transducer; (2) SCEE attenuating the backscattered signal by $>$50\% in 2 ME cycles; (3) PWM ME backscatter achieving 17.73 kbps data rate and 0.9 pJ/bit efficiency at 331 kHz carrier; (4) reliable operation at up to 5 cm distance with $<$8.5$\times10^{-5}$ BER using a lightweight multi-layer-perceptron (MLP) neural network for demodulation.
%


This article extends \cite{yu_millimetric_2024} and is organized as follows: Section II elaborates on the principles of SCEE and the modeling of ME transducers. Section III describes the design and optimization of the circuit and system. Section IV presents the experimental results, including chip measurement and in-vitro validations. Section V concludes this article.

\section{Principles of Switched-Capacitor Energy Extraction (SCEE)} \label{sec_principle}

\subsection{Basic Model of ME Transducer}

\begin{figure}[t!]
\centering
\includegraphics[scale = 0.24]{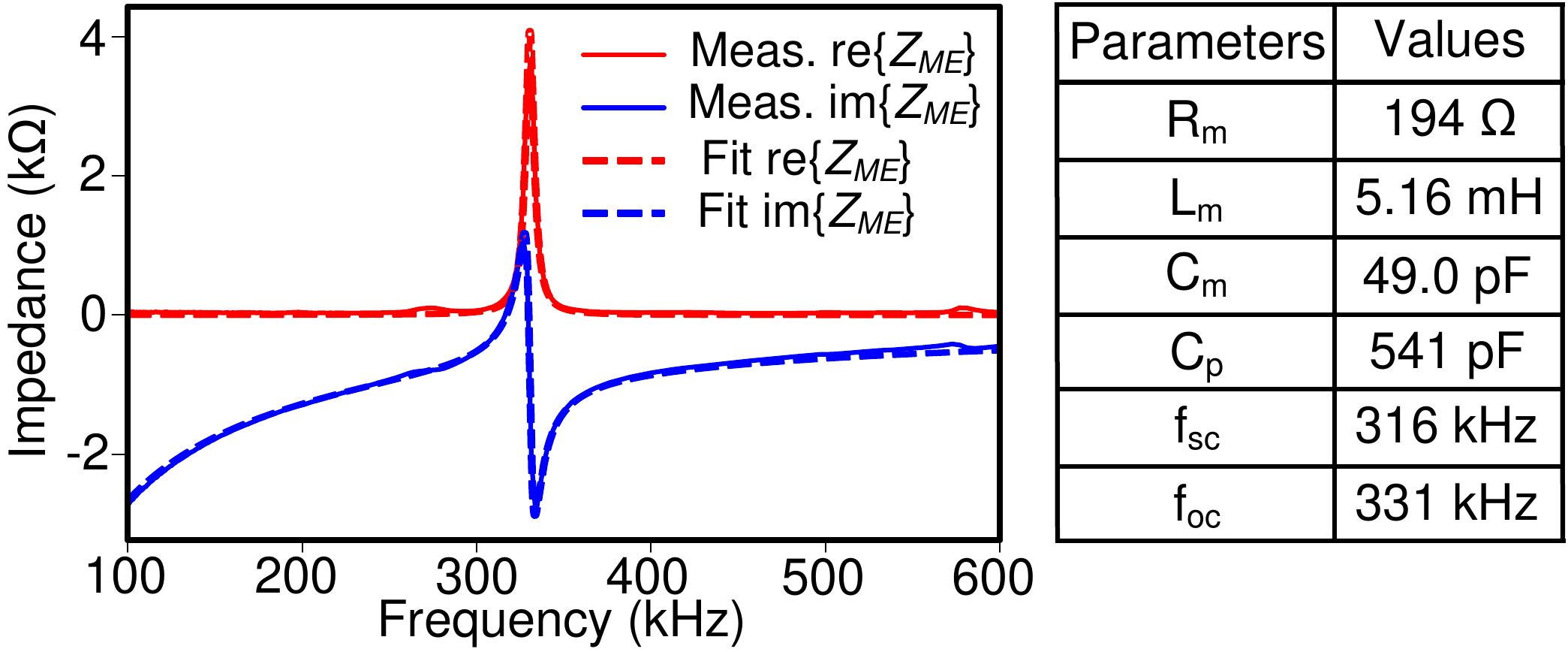}

\caption{Impedance fitting of an ME film and the corresponding model parameters, where $f_{sc}$ and $f_{oc}$ are the short-circuit and open-circuit resonance frequency calculated from the impedance model parameters.}
\label{1st_order_fitting}
\end{figure}

Magnetoelectric (ME) transducer is a composite laminate made with coupled magnetostrictive and piezoelectric films. The piezoelectric layer converts acoustic waves to electrical fields. 
Similar to modeling ultrasound transducers, we employ \textit{Van Dyke} circuit to model the electro-acoustic impedance~\cite{noauthor_ieee_1978}. 
Meanwhile, a transformer is used to represent the conversion from magnetic waves to acoustic waves \cite{alrashdan_wearable_2021}, as shown in Fig.~\ref{Original_model}. To obtain the parameters for this model, we first measure the output impedance of the fabricated ME films using a network analyzer. When the ME film is not driven by an alternating magnetic field, one can ignore the AC source in Fig.~\ref{Original_model} and obtain an impedance model shown in Fig.~\ref{Input_impedance}. The measured data is then fit into the impedance model. Fig.~\ref{1st_order_fitting} shows the measured and modeled impedance of a 5$\times$2 mm$^2$ ME transducer that we use in this work, along with the model fitting parameters. This model is subsequently utilized in circuit simulations to determine design parameters, such as the value of the flying capacitors in the SCEE interface.


\begin{figure}[t!]
\centering
\includegraphics[scale = 0.4]{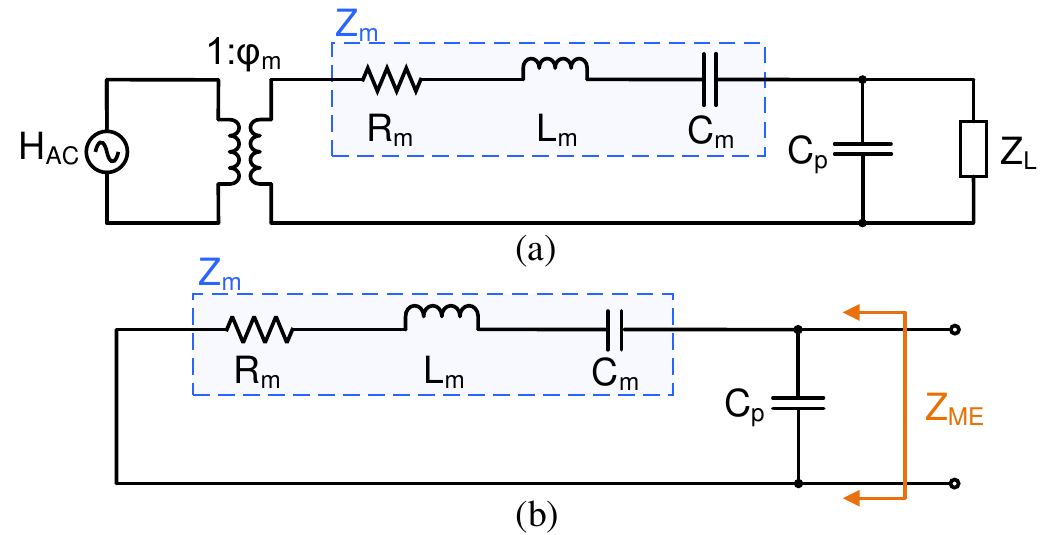}
\caption{(a) electrical and (b) simplified impedance model of a ME transducer.}
\label{Models}
\end{figure}

\subsection{Switched-Capacitor Energy Extraction}




\begin{figure}[t!]
\centering
\includegraphics[scale = 0.92]{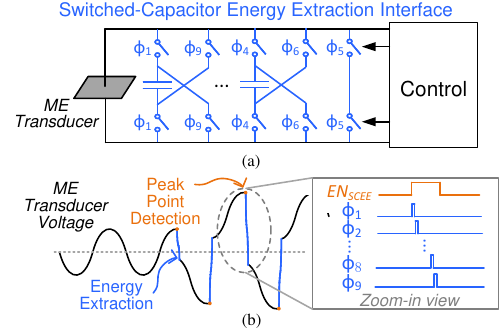}
\caption{(a) Diagram of switched-capacitor energy extraction interface and (b) its operating waveform.}
\label{principle}
\end{figure}

Our PWM ME backscatter scheme demands the ability to induce a detectable amplitude reduction in the transducer's output at an exact time during the ringdown phase. To achieve a high SNR, the amplitude reduction must be sharp, and the residual voltage must be minimal. Since the transducer voltage is induced by the acoustic vibration of the piezoelectric layer, reducing the amplitude necessitates extracting and dissipating the mechanical kinetic energy stored in the transducer. Simply shorting the two terminals of the ME transducer is ineffective, as it will cause the transducer to enter short circuit resonance, leaving a significant amount of energy in the resonator that continuously emits signals. 
In fact, this target aligns with the goal of piezoelectric energy harvesting (PEH) circuits, which is to maximize the extraction of kinetic energy from a piezoelectric transducer. Therefore, we drew inspiration from state-of-the-art PEH interface designs to develop the energy extraction circuit for ME backscatter telemetry.

Conventional PEH circuits, such as synchronized switch harvesting on inductor (SSHI) \cite{ramadass_efficient_2010, sanchez_parallel-sshi_2016} and synchronous electrical charge extraction (SECE) \cite{hehn_fully_2012, kwon_single-inductor_2014, dini_nanopower_2016 } designs, require bulky off-chip inductors to achieve good energy extraction performance. Moreover, zero-current detection in the switched inductor schemes will consume significantly more power with ME transducers that operate at much higher frequencies (hundreds of kHz) than typical PEH interfaces ($<$kHz). In contrast, the synchronized switch harvesting on capacitors (SSHC) technique achieves comparable performance by employing a series of capacitors to flip the voltage across the transducer synchronously~\cite{du_inductorless_2017}. Based on the principle of SSHC, the capacitor array should have comparable capacitance as the output capacitor of the transducer, which can be integrated on-chip in our case ($\sim$1 nF). Since SSHC is based on charge sharing, the requirements for on-resistance of the switches and timing precision are less stringent compared to that of the inductor counterpart, making it more suitable for our miniaturized ME implant. 

Fig.~\ref{principle} illustrates the operations of our switched-capacitor energy extraction (SCEE) method to quickly dissipate energy from an ME transducer and facilitate PWM backscatter. 
Similar to SSHC, SCEE employs sequentially activated switched capacitors to extract energy from the piezoelectric-layer capacitor $C_p$ at the peak points of the sinusoidal ME signal, where $C_p$ holds the most energy. Following energy extraction, $C_p$ is used to flip the transducer's voltage $V_{ME}$, effectively boosting the amplitude of $V_{ME}$ in the next half cycle, if $V_{ME}$ is not clamped by the rectifier's output. When SCEE boosts the transducer voltage, the output power of the ME transducer ($P_{ME} = V_{ME} \times I_{ME}$) increases, since the ME current remains unchanged due to the transducer's high quality factor. In this way, the mechanical energy in the ME transducer can be quickly extracted and dissipated (or transferred to the rectifier's output if clamped). 
The voltage flipping process repeats eight times over four ME cycles to ensure sufficient energy extraction and amplitude reduction while preserving a good timing resolution for PWM. 
Four flying capacitors with a total on-chip capacitance of 1.2 nF are used for the SCEE interface considering the trade-offs among hardware complexity, on-chip area, and energy extraction performance.

\begin{figure}[t!]
\centering
\includegraphics[scale = 0.24]{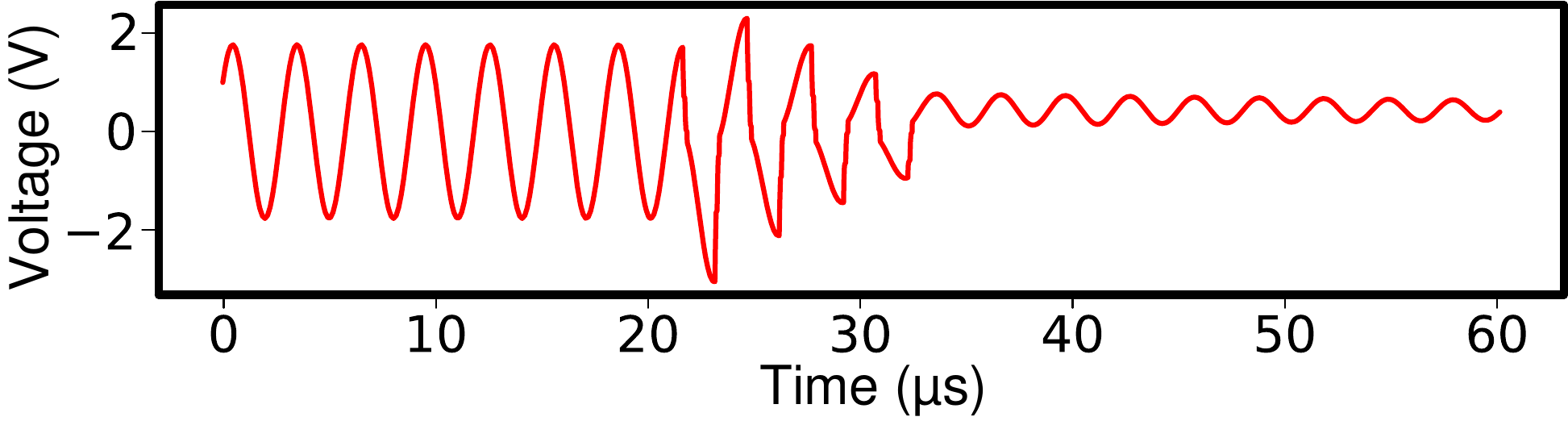}

\caption{Simulated ME voltage with SCEE using the basic circuit model.}
\label{VAC_1st}
\end{figure}

Using the model in Fig.~\ref{Input_impedance}, we simulated the SCEE interface circuitry performing energy extraction 8 times when the external field is turned off. For simplicity, no rectifier is connected in these simulations, so there is no voltage clamping. Fig.~\ref{VAC_1st} shows the simulated ME transducer voltage $V_{AC}$, where the voltage flipping points are well aligned with the peaks and the amplitude of ringing is drastically decreased. 

\begin{figure}[t!]
\centering
\includegraphics[scale = 0.24]{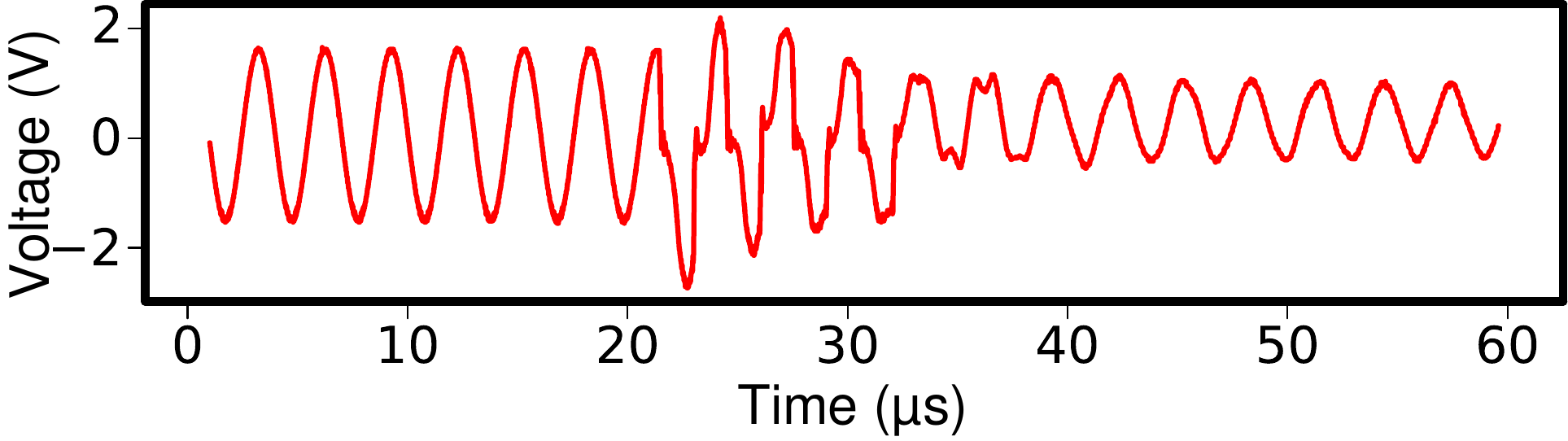}

\caption{Measured $V_{AC}$ waveform with SCEE interface.}
\label{VAC_meas}
\end{figure}

\begin{figure}[t!]
\centering
\includegraphics[scale = 0.4]{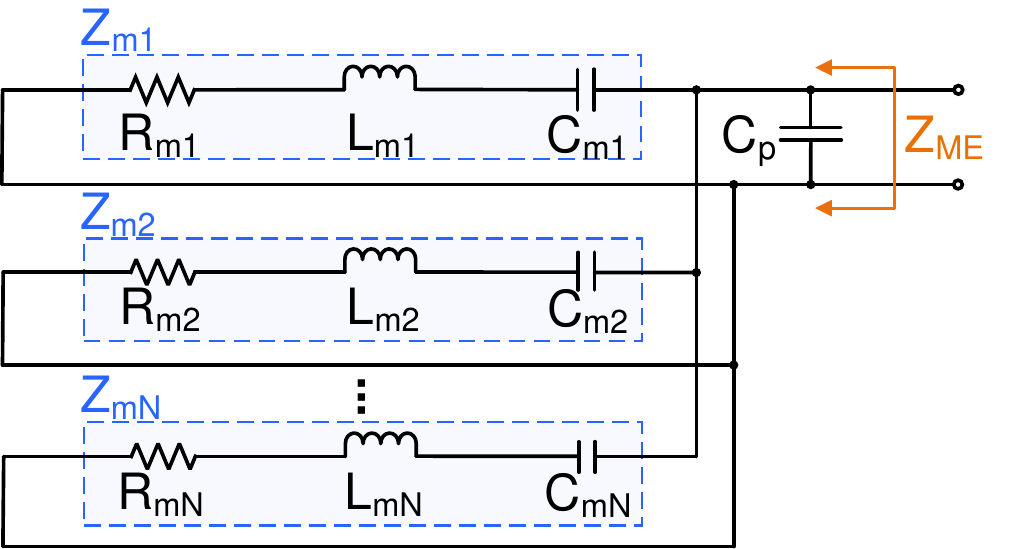}

\caption{High-order ME impedance model that consists of N vibration modes.}
\label{high_order_impedance}
\end{figure}

\begin{figure}[t!]
\centering
\includegraphics[scale = 0.23 ]{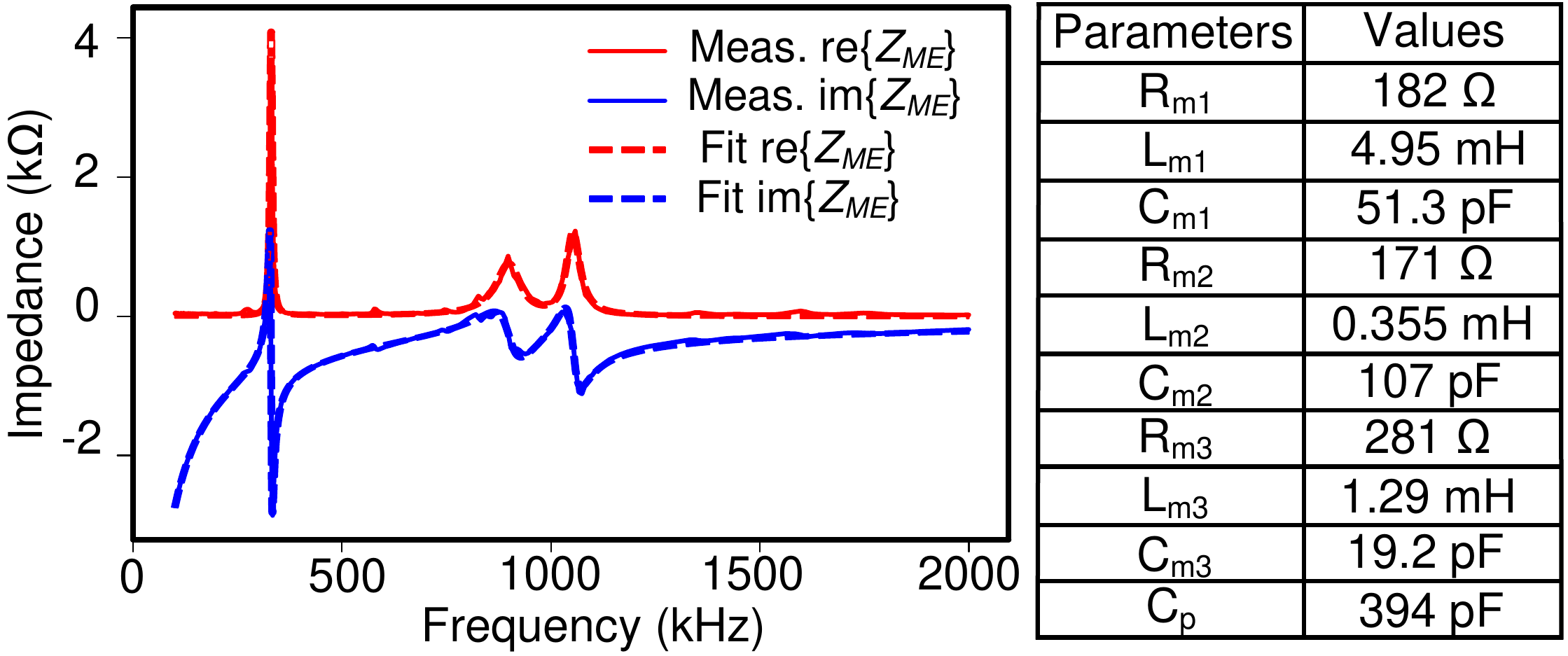}

\caption{Impedance measurement and fitting up to 2 MHz and the corresponding parameters of the 3rd order impedance model.}
\label{3rd_order_fitting}
\end{figure}

\begin{figure}[t!]
\centering
\includegraphics[scale = 0.24]{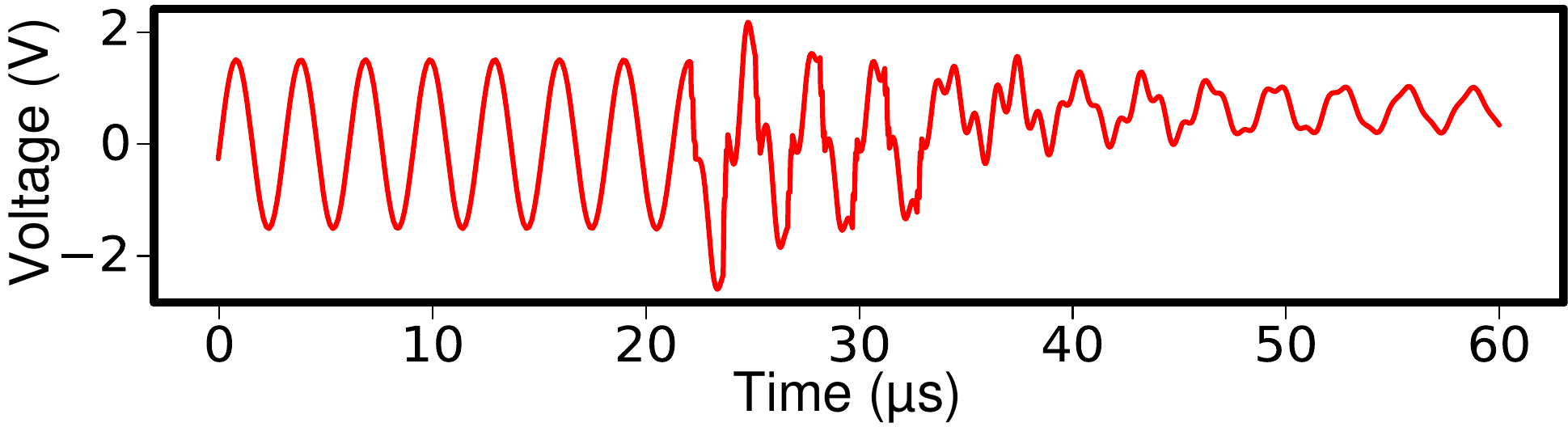}

\caption{Simulated $V_{AC}$ waveform with SCEE using the high-order model.}
\label{VAC_3rd}
\end{figure}

\subsection{High-Order Model of ME Transducer}

\begin{figure*}[t!]
\centering
\includegraphics[scale = 0.75]{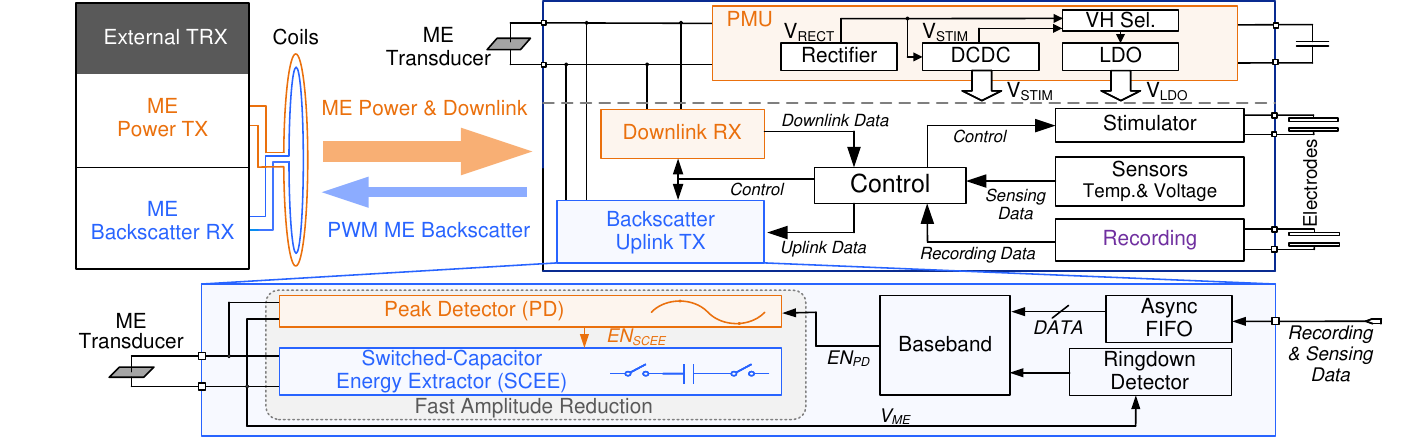}

\caption{System diagram of the ME implant SoC.}
\label{system_diagram}
\end{figure*}

While using a PCB-level SCEE prototype to validate the concept, we observe noticeable deviations from the simulation results obtained with the basic ME model, as shown in Fig.~\ref{VAC_meas}. 
After voltage flipping, the transducer voltage amplitude first decreases and then increases in the opposite direction. Moreover, frequency components that are higher than the carrier frequency appear after voltage flipping and persist several more cycles after completing the SCEE operations. 
The peak points become difficult to distinguish, and the residual voltage after SCEE is higher than the baseline. 

Based on the observation of high-frequency components, we attempt a higher-order model to emulate the ME transducer's behavior. Prior works have also revealed the second vibration mode of an ME transducer and utilized it for uplink communication \cite{hosur_magsonic_2023}. However, there is no circuit model in the literature that describes the other vibration modes. Here, we suggest a high-order ME model to capture the effects of high-frequency vibration modes in energy extraction and can potentially extend to other designs. 
The primary vibration mode of an ME film is the length mode, which operates at a low frequency of hundreds of kilohertz. However, the ME film also exhibits piezoelectric effects in other directions, such as width and thickness. These modes have higher frequencies and generally contribute minimally to the ME transducer voltage when driven by a magnetic field, as they are far off-resonance. During energy extraction, however, the interface circuit induces sudden drops in ME voltage, which not only extracts energy from the main resonance mode but also excites other resonance modes due to the wide bandwidth of the pulse-shaped waveforms. Consequently, these extra resonance modes affect the energy extraction process.

Fig.~\ref{high_order_impedance} shows our high-order ME impedance model. Here, each resonance is represented by an impedance $Z_{m}$, where the $R_m$, $L_m$, and $C_m$ correspond to damping, mass, and stiffness in the mechanical domain. $C_p$ is the lumped electrical parasitic capacitor between the two ME terminals. 
Using this model, we can analyze the ME transducer's behavior in SCEE through simulation tools. Note that this model does not include input sources in each branch, as the external TX is turned off during SCEE. However, if we need to examine the behavior when the ME film is powered by a magnetic wave, we can introduce the sinusoidal source and transformer to the main resonance branch while treating the other branches as loads.

Fig.~\ref{3rd_order_fitting} displays the impedance measurement and model fitting results for the same ME film measured in Fig.~\ref{1st_order_fitting}, but covering a higher frequency range up to 2 MHz. 
The impedance curves clearly exhibit three dominant resonance peaks, so a 3rd-order impedance model was selected for fitting. The fitting results align well with the impedance measurement across the entire frequency range.

We further simulated the SECC interface circuit with the 3rd-order ME model to study its impacts on energy extraction. The simulated $V_{AC}$ waveform in Fig.~\ref{VAC_3rd} nicely mimics the measured one in Fig.~\ref{VAC_meas} with two prominent features. 
First, the $V_{AC}$ voltage shows a reversion right after each voltage flipping. Second, when the whole SCEE process is done, the $V_{AC}$ waveform is still distorted for a few cycles, which is more intense in the simulation. These features are induced by the higher-frequency resonance excited by the SCEE operation, which justifies our high-order ME model. 

Despite the non-ideal behavior caused by the high-frequency resonance modes, SCEE still effectively reduces the voltage amplitude in the ME transducer. According to the simulation results in Fig.~\ref{VAC_1st} and Fig.~\ref{VAC_3rd}, the amplitude reduction effect only degrades by 8.9\% (from 82.4\% to 73.5\%) when using the high-order model. This slight decline does not significantly impact the performance of the PWM backscatter and thus does not warrant extra efforts for mitigation. The high-order effects and high-quality ME backscatter are further demonstrated with our integrated SoC, as discussed in Section \ref{measurement}.

\section{System and Circuit Implementation}

\subsection{System Overview}

\begin{figure}[t!]
\centering
\includegraphics[width=\linewidth]{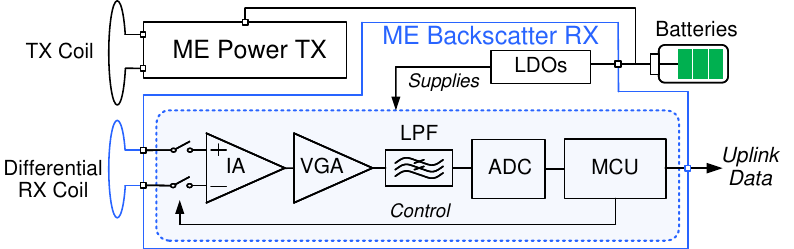}

\caption{Diagrams of the portable external TRX.}
\label{RX_board}
\end{figure}

Fig.~\ref{system_diagram} presents the system diagram of the ME implant's fully integrated SoC, which integrates functions for bio-stimulation, neural recording, temperature sensing, and bidirectional communication. During operation, the input ME power is rectified and regulated by the power management module to generate a stable 1 V supply voltage. A DCDC converter \cite{yu_wireless_2022} is used to generate the stimulation voltage. The LDO’s supply is connected to the higher voltage between $V_{RECT}$ and $V_{STIM}$ to guarantee enough voltage headroom. To ensure robust downlink and uplink operation, a decoupling capacitor of 500 nF is placed at the rectifier’s output to ensure a stable $V_{RECT}$ during the ringdown phase. 
Simultaneously, the downlink data is recovered by the downlink receiver (RX) and sent to a central controller. This downlink data contains stimulator parameters and other SoC configurations.
When the SoC is in uplink mode, the controller encodes data from sensors or the recording front-end and transmits it via the backscatter uplink transmitter (TX). This transmission is enabled by the SCEE interface and the peak detector, which allow for rapid amplitude reduction at specific moments. An asynchronous FIFO serves as a buffer for handling any mismatch between uplink and recording data rates. In the uplink phase, the power TX needs to be turned off frequently to create the ringdown phase, during which the power transfer is interrupted and the received power is halved approximately. However, thanks to the low power consumption of the uplink telemetry and chip, the delivered power is enough to sustain the system’s operation.
Meanwhile, as shown in Fig.~\ref{RX_board}, we developed a custom transceiver (TRX) using off-the-shelf components to transfer power and downlink data to the implant, and to receive and decode the uplink data via PWM ME backscatter.

\begin{figure}[t!]
\centering
\includegraphics[scale = 0.85]{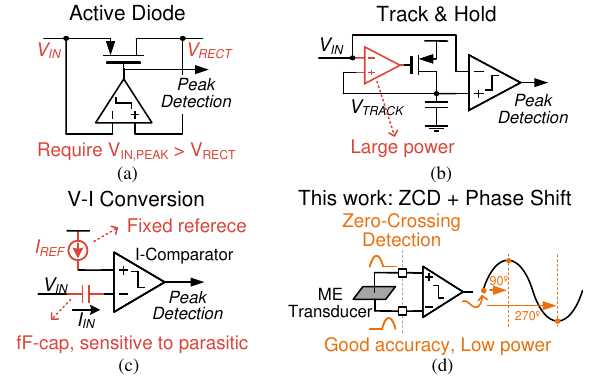}
\caption{Peak detector designs based on (a) active diode~\cite{sanchez_parallel-sshi_2016, yue_resonant_2024}, (b) track and hold~\cite{dini_nanopower_2016}, (c) voltage-to-current conversion~\cite{hehn_fully_2012, kwon_single-inductor_2014}, and (d) zero-crossing-detection with phase shift.}
\label{PD_principle}
\end{figure}

\begin{figure}[t!]
\centering
\includegraphics[width=\linewidth]{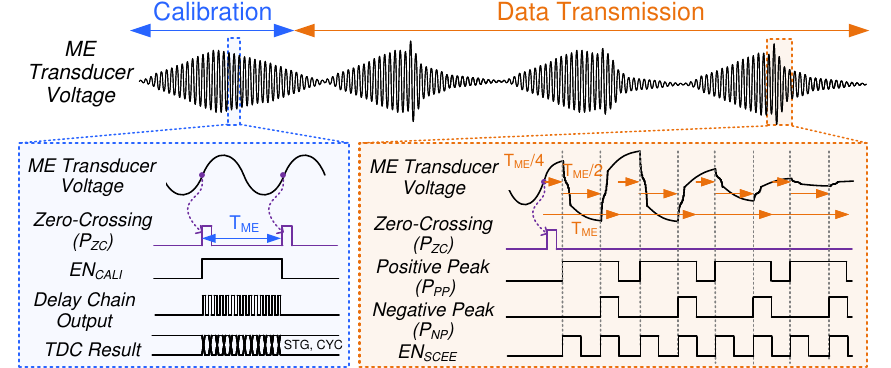}

\caption{Operating waveforms of the proposed peak detector.}
\label{PD_phase}
\end{figure}

\subsection{Peak Detector (PD)}

An accurate, low-power PD is essential for achieving the expected performance and efficiency of SCEE and PWM ME backscatter. Fig.~\ref{PD_principle} summarizes conventional peak detectors in prior PEH works. In \cite{sanchez_parallel-sshi_2016, yue_resonant_2024}, the comparator output in the active diode in the rectifier is used to determine the peaking, as shown in Fig.~\ref{active_diode}. This method is simple but necessitates an input voltage larger than the rectifier's output $V_{rect}$, which is typically the case in SSHI and SSHC designs. However, in the ME backscatter uplink, the modulation occurs in the ringdown, during which the input voltage is decaying and smaller than $V_{rect}$, rendering this method unusable. 
In \cite{dini_nanopower_2016}, a track \& hold PD is used for SECE,
which doesn't have the $V_{rect}$ limitation, shown in Fig.~\ref{Track_hold}. But since the ME transducer works at a much higher frequency (331 kHz in this work) than in typical PEH circuits ($<$1~kHz), a wide-bandwidth amplifier is required to ensure a short delay between $V_{TRACK}$ and $V_{IN}$, which burns much power. Fig.~\ref{VI_conversion} shows the voltage-to-current-conversion-based PD used in \cite{hehn_fully_2012, kwon_single-inductor_2014}. It demands a series capacitor and a current reference to transform the input voltage into a current. When the peaks arrive, the current becomes zero and triggers the comparator, as shown in Fig~\ref{VI_conversion}. However, the capacitor nodes and current reference are sensitive to parasitic and input variation. 

To meet the requirements, we developed a specialized high-performance low-power PD combining zero-crossing detection and phase shifting, as shown in Fig.~\ref{ZCD_PS}, by exploiting the fact that detecting zero points is easier to implement and more accurate, and peak points are always \ang{90} and \ang{270} away from the zero points in a sine wave.
Fig.~\ref{PD_phase} illustrates the operation of the peak detector in two phases: calibration and data transmission. At the beginning of each uplink session, a calibration phase is performed to convert ME's period $T_{ME}$ to digital codes. 
The calibration is done during the ringdown phase to avoid inaccurate timing caused by the mismatch between the TX driving frequency and ME film's self-resonance frequency.
After calibration, TX enters data transmission mode, and PD produces peak-aligned pulse trains ($EN_{SCEE}$) for controlling the SCEE. Here, a digital phase shifter (DPS) in the PD adds $T_{ME}$/4 and 3$T_{ME}$/4 delays to the output pulse of the zero-crossing detector (ZCD) to locate positive and negative peaks. The baseband controller tracks the number of ringdowns and ME clock over the entire uplink session, and activates PD for 4 ME cycles at specific moments based on uplink data. Within these 4 cycles, ZCD is activated only in the first cycle while DPS creates 8 $EN_{SCEE}$ pulses based on its calibrated delay line, lowering the average power to 15 nW.

In the PD, $T_{ME}$/4 and 3$T_{ME}$/4 delays are required to derive the peak points from the zero-crossing point. Conventionally, shift register and multiplier are necessary to calculate the result, which is area and power-consuming, especially considering the 12-bit TDC output. In this work, the DPS employs a shift-only arithmetic to approximate the $T_{ME}$/4 and 3$T_{ME}$/4 delays with minimum overheads.

\begin{figure}[t!]
\centering
\includegraphics[scale = 0.63]{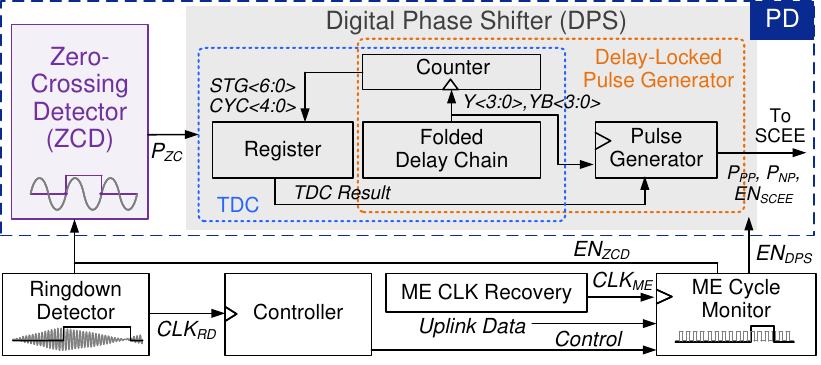}

\caption{Block diagram of the peak detector (PD).}
\label{PD_block_diagram}
\end{figure}

\begin{figure}[t!]
\centering
\includegraphics[width=\linewidth]{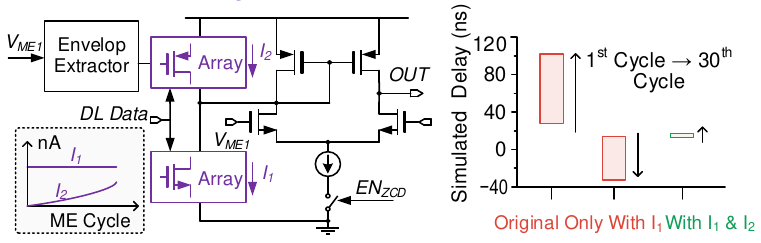}

\caption{Schematic of the adaptive delay-compensated zero-crossing detector and simulated delay drift from the 1$^{st}$ to the 30$^{th}$ cycle during the ringdown.}
\label{ZCD_circuit}
\end{figure}

Fig.~\ref{PD_block_diagram} shows the block diagram of the peak detector and related modules. When a ringdown is detected during the uplink phase, the uplink controller will determine when to modulate the ME transducer voltage based on the uplink data. The $EN_{ZCD}$ and $EN_{DPS}$ signals are sent to the peak detector to enable the zero-crossing detector and digital phase shifter. When the ZCD detects the zero-crossing point $P_{ZC}$, the DPS will shift the signal by $T_{ME}$/4 and 3$T_{ME}$/4 to find the peak points to do the voltage flipping using the folded delay chain.

\subsection{Zero-Crossing Detector (ZCD)}

Fig.~\ref{ZCD_circuit} illustrates the schematic of the ZCD in this design. The ZCD is enabled at a specific cycle during every ringdown in the uplink phase, according to the uplink data. Its accuracy directly affects the efficacy of the SCEE.
For accurate detection, the ZCD uses a comparator with an intentionally added offset current, which minimizes detection delay by intentionally adding an offset without significantly increasing the biasing current. However, using a fixed sinking current $I_1$ leads to overcompensation during later ME ringdown cycles when the signal amplitude is smaller. To address this issue, we introduced an adaptive sourcing current $I_2$ that increases with each cycle to complement $I_1$. Simulation results demonstrate that incorporating the adaptive current greatly reduces the ZCD delay and ensures more consistent performance throughout the entire ringdown phase. $I_2$ is biased by an ME envelope extractor that is also necessary for downlink, thus eliminating the need for additional hardware.

\subsection{Digital Phase Shifter (DPS)}

\begin{figure}[t!]
\centering
\includegraphics[width=\linewidth]{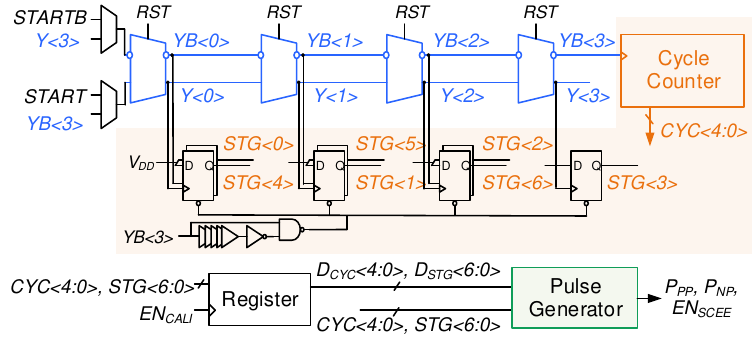}

\caption{Schematic of the digital phase shifter (DPS).}
\label{DPS}
\end{figure}

The DPS comprises a differential 4-stage folded delay chain with a 15 ns step (i.e., 0.5\% of $T_{ME}$), a cycle counter, sampling DFFs, and a digital pulse generator, as shown in Fig.~\ref{DPS}. The folded delay chain is used to mitigate the exponentially increasing area requirements for higher time resolutions \cite{chen_dct-ram_2022}. During the calibration phase, the DPS works as a TDC to record the time interval between two zero-crossing points. The output of the TDC comprises two parts: the cycle counter's output ($D_{CYC}\langle4:0\rangle$, binary) and the stage counter's output ($D_{STG}\langle6:0\rangle$, thermometer). The digital code is stored in the sampling DFFs. At the data transmission phase, the DPS is triggered when the cycle counter's output matches the desired delay, and pulsing signals will be sent to the SCEE interface to execute the modulation. 

As mentioned before, shift-only arithmetic is employed to derive the $T_{ME}$/4 and 3$T_{ME}$/4 delay from the $T_{ME}$'s TDC output code, which saves considerable computation overhead without sacrificing much accuracy. The digital code of $T_{ME}$ can be represented by
\begin{equation}
\resizebox{0.95\columnwidth}{!}{
$ T_{ME} = t_{delay}\times[8\times dec\{D_{CYC}\langle 4:0\rangle\} + dec\{D_{STG}\langle6:0\rangle\}]  $
}
\label{eq_1}
\end{equation}
where $dec\{D\}$ represents the decimal format of the digital code $D$. When divided by 4, the first half of \eqref{eq_1} can be shifted right by 2 bits, resulting in $8\times dec\{D_{CYC}\langle4:2\rangle\}$ plus $2\times \{D_{CYC}\langle1:0\rangle\} $. The second half, which consists of the thermometer part from the stage counter, can be approximated by $D_{STG}\langle3\rangle$ when divided by 4. This means that when $D_{STG}\langle6:0\rangle$ equals or is larger than 0001111 (thermometer), the division result is set to~1; when it is smaller, the result is set to~0. Therefore the final $T_{ME}$/4 can be expressed as
\begin{equation}
\resizebox{0.8\columnwidth}{!}{
$\begin{aligned}
 T_{ME}/4 =& t_{delay} \times [8\times dec\{D_{CYC}\langle 4:2\rangle\} \\
&+ dec\{D_{CYC}\langle1\rangle,D_{CYC}\langle0\rangle, D_{STG}\langle3\rangle \}] 
\end{aligned}$
\label{eq_2}
}
\end{equation}
where $\{D_{CYC}\langle1\rangle,D_{CYC}\langle0\rangle, D_{STG}\langle3\rangle \}$ is in a 3-bit binary format. The maximum error from this approximation is obtained when $D_{STG}\langle6:0\rangle$ equals 0000111 or 1111111, which leads to an error of $3 t_{delay} /4$.

The calculation for 3$T_{ME}$/4 is similar to \eqref{eq_2}, 
\begin{equation}
\resizebox{0.8\columnwidth}{!}{
$\begin{aligned}
 3T_{ME}/4 =& t_{delay} \times [8\times dec\{D_{CYC}\langle 4:1\rangle\} \\
&+ dec\{D_{CYC}\langle0\rangle,D_{STG}\langle5\rangle, D_{STG}\langle1\rangle \}] .
\end{aligned}$
\label{eq_3}
}
\end{equation}

According to \eqref{eq_2} and \eqref{eq_3}, the shift-only arithmetic of the DPS has an approximation error bounded to 3/4 of the single-stage delay, which leads to negligible impacts on the performance of switched-capacitor energy extraction while greatly simplifying the hardware.

\subsection{Recording Analog Front-End}


\begin{figure}[t!]
\centering
\includegraphics[scale = 0.58]{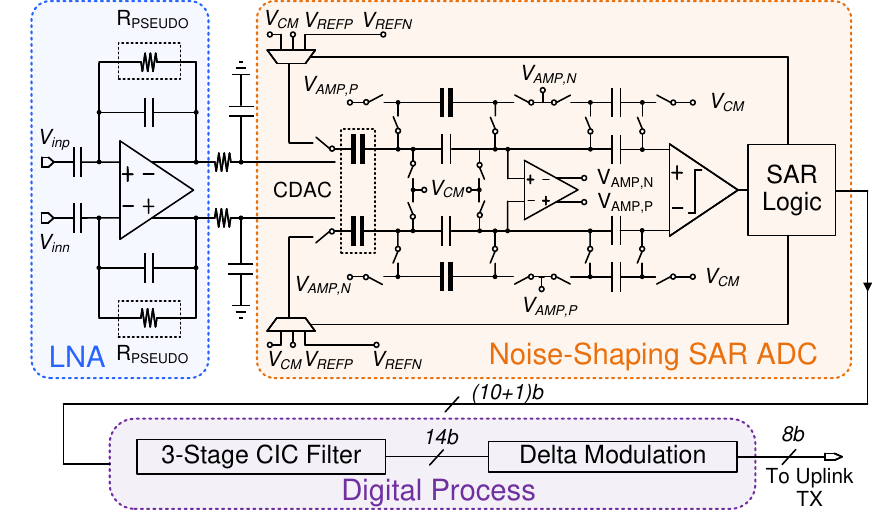}

\caption{Schematic of the recording AFE.}
\label{AFE}
\end{figure}

To demonstrate real-time recording, the prototype implant’s SoC includes a recording analog front-end (AFE) and supports continuous data streaming. Fig.~\ref{AFE} illustrates the implementation of the recording AFE. To expand the input dynamic range ($>$ 60 dB) and eliminate the need for a high-resolution ADC (ENOB $>$ 15 bits), we choose to implement the first stage of the AFE as a moderate-gain (gain = 15), current-reusing two-stage low noise amplifier (LNA). Because of its moderate gain, the neural signal is less likely to saturate the amplifier. Since the LNA drives the ADC directly, we implement a highly energy-efficient third-order noise-shaping SAR ADC \cite{wang_138-enob_2021}, which features a small sampling capacitor to relax the driving strength requirement of the LNA. The ADC has a 1 kHz bandwidth and an oversampling ratio (OSR) of 8. Due to the low OSR, the subsequent decimation filter is a simple 3-stage cascaded integrator-comb (CIC) filter. Since the amplitude of the bio-signal is relatively small and the frequency of the artifacts is low compared to the ADC’s bandwidth, a delta modulator is added to further reduce the final output to a 2~kSa/s 8-bit data stream for wireless uplink.

\section{Measurement Results}\label{measurement}

We prototyped the ME implant SoC in TSMC 180 nm CMOS technology. Fig.~\ref{Die} shows the micrograph of the 2.6 $mm^2$ SoC. The SoC was packaged with a 5$\times$2 mm$^2$ ME transducer into a 6.7 mm$^3$ implant, as shown in Fig.~\ref{implant_photo}. 


\begin{figure}[!t]
\centering
\includegraphics[width=.95\linewidth]{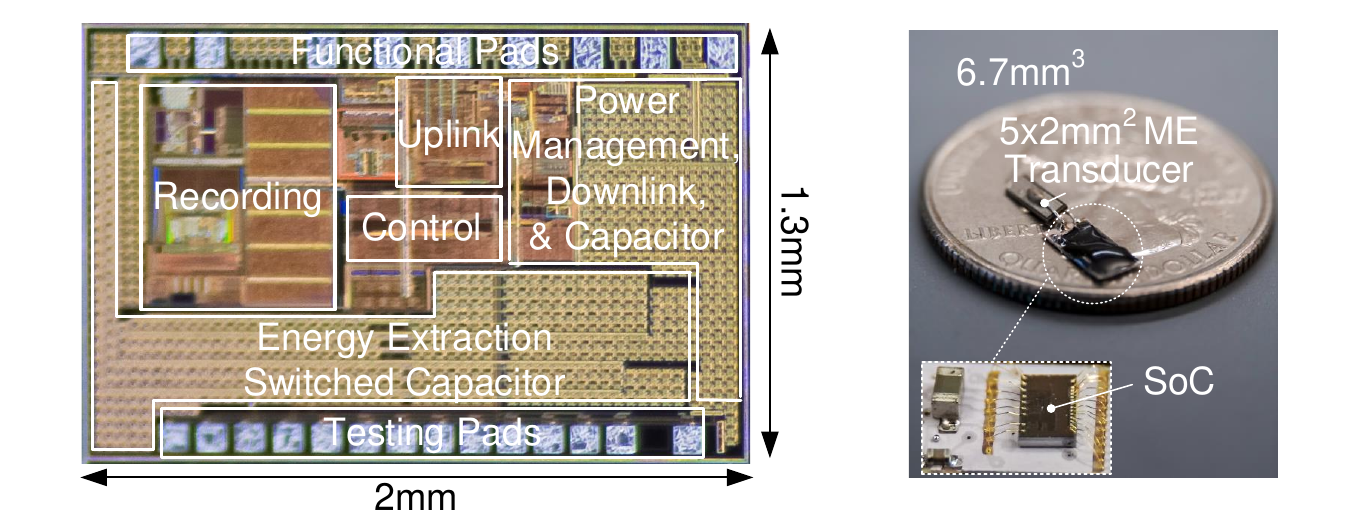}
\caption{Micrograph of the implant SoC and the device assembly.}

\label{photos}
\end{figure}

\subsection{System Validation}

\begin{figure}[t!]
\centering
\includegraphics[scale = 0.67]{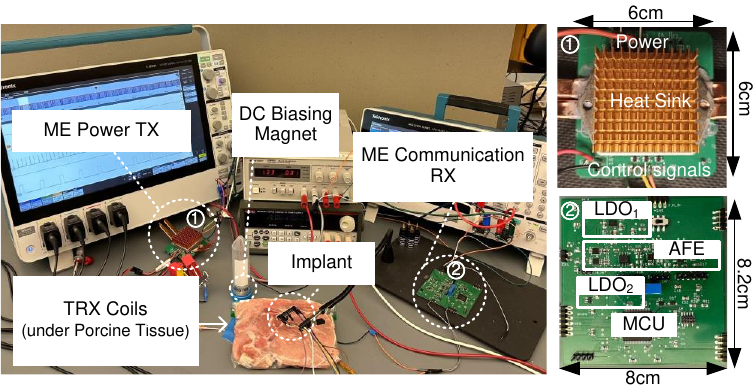}

\caption{\textit{Ex vivo} testing setup, including the TX and RX hardware.}
\label{Testing_setup}
\end{figure}

\begin{figure}[t!]
\centering
\includegraphics[width=\linewidth]{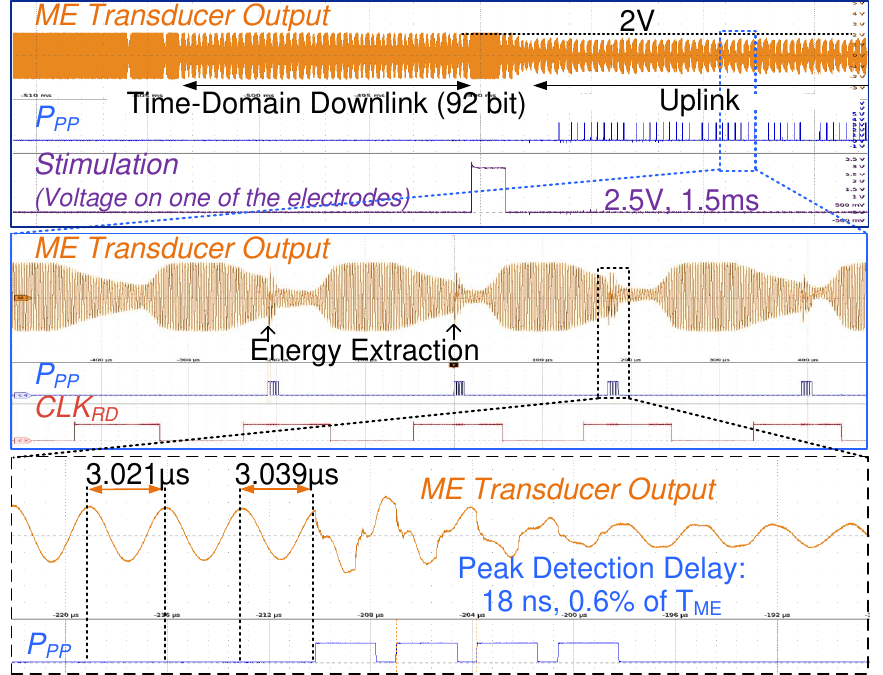}

\caption{Implant SoC operation waveform, with zoom-in views of the ME energy extraction process.}
\label{Trans_waveform}
\end{figure}

We first evaluated the functionalities of the full system \textit{ex vivo} by placing a 2 cm-thick porcine tissue between the implant and the external TRX. The testing setup is displayed in Fig.~\ref{Testing_setup}. The TX is programmed to drive the TX coil with targeted ME frequency and patterns to deliver power and downlink data to the implant, while the RX collects the backscattered signal and decodes the uplink data. The TX and RX coils are assembled in a concentric fashion to cancel the TX interference. The DC biasing magnet required for ME operation is placed near the implant. Fig.~\ref{Trans_waveform} shows the overall operation waveforms of the implant. Specifically, the implant is magnetoelectrically powered and programmed with time-domain modulated downlink data. After programming, the SoC performs different tasks, including biphasic stimulation and uplink telemetry. The zoom-in views show the ME transducer voltage during the PWM backscatter uplink. The SCEE extracts the ME transducer's energy at different time points during the ringdown based on PWM data. The prototype demonstrates energy extraction from the ME transducer near the optimal peak points, causing an immediate biasing flip. Despite the distortion in the waveform as discussed in Section \ref{sec_principle}, the 4-cycle energy extraction quickly dissipates the energy in the ME transducer, resulting in a more than 1 V amplitude decrease in the transducer's output voltage. The first peak detection pulse shows an error of 18 ns (i.e., 0.6\% of $T_{ME}$), demonstrating sufficient accuracy of the PD.

\begin{figure}[t!]
\centering
\includegraphics[width=.95\linewidth]{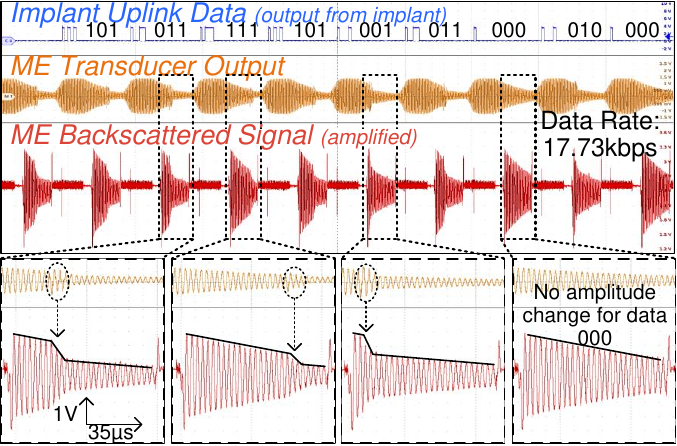}

\caption{Waveforms of the data decoding of the PWM ME backscatter.}
\label{Backscatter}
\end{figure}

\begin{figure}[t!]
\centering
\includegraphics[width=.95\linewidth]{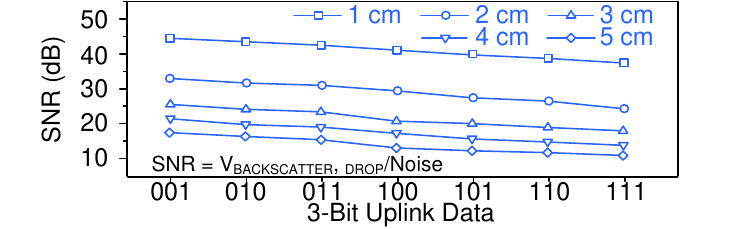}

\caption{Measured SNR of backscattered signal across different distances.}
\label{SNR}
\end{figure}

\subsection{PWM Backscatter Testing}

Next, we evaluated the data rate and SNR of the backscattered signal. 
Fig.~\ref{Backscatter} offers a sample measurement of the signal after RX AFE, with clearly distinguishable pulse widths during ringdown, modulated by the 3-bit uplink data. Here, each 3-bit data packet takes a 24-ME-cycle excitation phase, succeeded by a 32-ME-cycle ringdown phase, resulting in a data rate of 17.73 kbps with a 331 kHz carrier. To trade-off between the SNR and the distinction between codes, 3-ME-cycle time interval is selected for the pulse-width modulation.

Besides, Fig.~\ref{SNR} depicts the SNR results across different TRX-implant distances. In our setup, SNR is defined as the voltage drop in the backscattered signal divided by the noise. Owing to fast amplitude reductions at accurate time intervals, the received signal achieves an SNR $>$10.9 dB at up to 5 cm distance. The SNR varies with different uplink data because of our PWM encoding methods. For example, the data pattern "001" is modulated at the beginning of the ringdown, where the ME transducer has the highest remaining voltage, so it has the highest voltage drop and SNR.

\begin{figure}[t!]
\centering
\includegraphics[scale = 0.9]{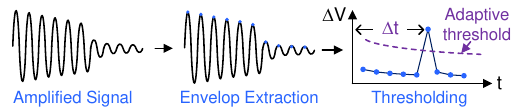}

\caption{Illustration of time-domain drop detection-based demodulation.}
\label{Demodulation_drop}
\end{figure}

\begin{figure}[t!]
\centering
\includegraphics[width=.95\linewidth]{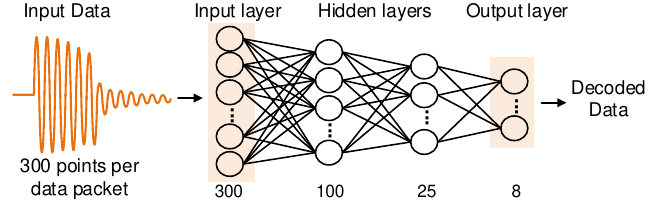}

\caption{Illustration of the MLP model for demodulation.}
\label{Demodulation_MLP}
\end{figure}

\begin{figure}[t!]
\includegraphics[width=.95\linewidth]{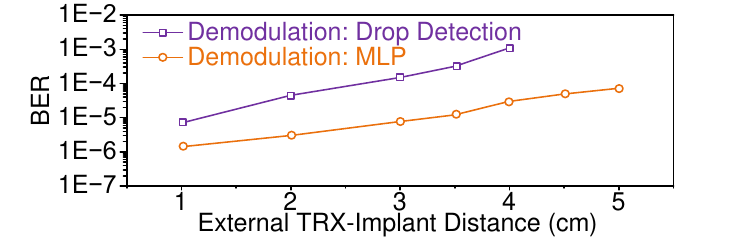}

\caption{Measured bit error rate (BER) of uplink communication.}
\label{BER}
\end{figure}

\subsection{Demodulation of Backscattered Signal}

We present two algorithms to decode uplink data from the backscattered signal: straightforward time-domain drop detection, and a compact 4-layer Multi-Layer Perceptron (MLP) neural network. The backscattered signal after the RX front-end is sampled by a 12-bit ADC at 2 MS/s. 

Fig.~\ref{Demodulation_drop} illustrates the principles of the time-domain drop detection method. First, the envelope of the amplified backscattered signal is extracted by identifying the peaks in the waveform. Next,  the voltage differences between adjacent peaks are compared with an adaptive threshold that decays over time. When the $\Delta V$ signal exceeds the threshold, it triggers the decoder to read the corresponding timestamp and generate the output data based on a look-up table (LUT). This method is straightforward, featuring low latency and low power consumption. However, as the distance between the implant and RX becomes larger, the waveform is susceptible to noise, interference, and body movement, causing a large variation in the $\Delta V$ signal and eventually reducing the decoding accuracy.

To enhance robustness against low SNR signals, we trained a 4-layer MLP using PyTorch. The network comprises four layers with 300, 100, 25, and 8 fully connected neurons, respectively, using ReLU as the activation function. All weights and activations are quantized to 5 bits. The training dataset includes backscattered signals recorded from 1 cm to 5 cm. Each training sample is segmented by the trigger signal and consists of a 150-$\mu$s waveform with 300 data points. During inference, the time-domain signal is cropped based on the data packet and processed by the MLP, which classifies the waveform into one of eight output codes. The quantized model uses only 20 KB of memory, making it feasible for inference on an MCU with low cost and low power consumption.

Fig.~\ref{BER} shows the bit error rate (BER) results with drop detection and MLP decoding. The drop detection provides sufficient BER (\textless 1$\times$10\textsuperscript{-5}) within a shorter distance while the MLP decoding offers higher accuracy across a considerable distance, yielding a BER of 8.5$\times 10^{-5}$ at 5 cm.

\subsection{Demonstration of Continuous Wireless LFP Recording}

\begin{figure}[t!]
\centering
\includegraphics[width=\linewidth]{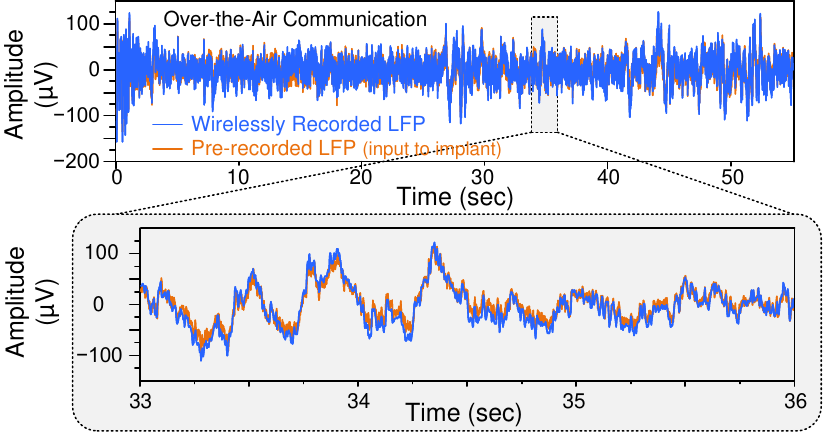}

\caption{Waveform of wirelessly recorded LFP signal versus ground truth.}
\label{Recording_waveform}
\end{figure}


\begin{figure}[!t]
\centering
\includegraphics[width=.9\linewidth]{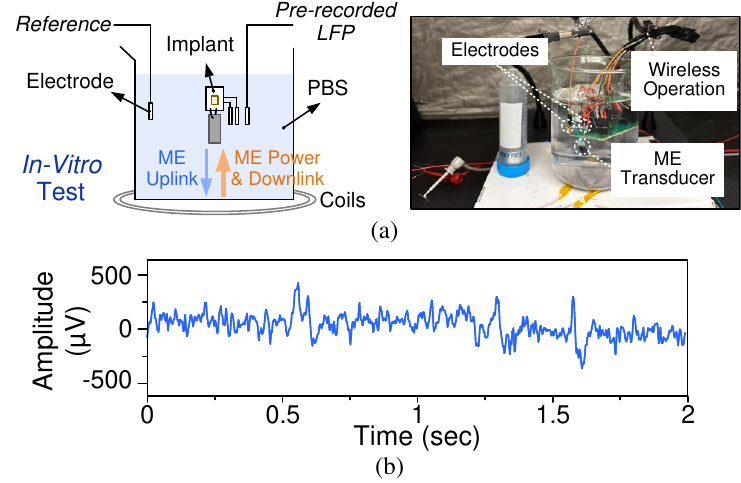}

\caption{(a) Measurement setup for the \textit{in vitro} test and (b) the corresponding recorded LFP signal.}

\label{Demodulation}
\end{figure}

We demonstrate the system's wireless recording functionality in two stages. First, pre-recorded local field potential (LFP) waveforms from rats are fed to the device through wires, while the recorded data is wirelessly streamed to the RX. The received signal via PWM ME backscatter closely matches the ground truth, as shown in Fig.~\ref{Recording_waveform}. In the second stage, we \textit{in vitro} validated an encapsulated untethered implant in phosphate-buffered saline (PBS). Fig.~\ref{Invitro_setup} shows the diagram and a picture of the testing setup \textit{in vitro}. The TRX coils are placed under the beaker for power transfer and bidirectional communication. The pre-recorded LFP signal is delivered through a pair of electrodes immersed in PBS. The working electrode is placed close to the implant, while the reference electrode is placed further away on the other side so that the chip can receive the differential signal. Fig.~\ref{In_vitro_data} shows the LFP recording result. Ground truth data is not provided since the signal received by the implant is highly dependent on the placement of the electrodes and implants. 

\subsection{System Power Consumption}

The presented wireless ME SoC consumes a total power of 74.8 $\mu$W, with a power transfer efficiency of 0.37\%. About 80\% of this power is used by the power management and control blocks. The uplink communication module consumes a minimal power of 0.45 $\mu$W, accounting for only 0.6\% of the total power, thanks to the backscatter scheme and highly duty-cycled circuit design. The uplink communication bit energy, defined as the uplink power divided by data rate, is 0.9 pJ/bit. The low bit energy is achieved thanks to the highly duty-cycled peak detector, which consumes the most power of the uplink module. The bit energy remains unchanged when distance varies because the uplink power is insensitive to the distance.

\subsection{Performance Summary and Comparison}

Table \ref{table} offers a comparison with state-of-the-art mm-scale wireless implants with uplink telemetry. The proposed ME system demonstrates a high efficiency of 0.37\% at 2 cm, outperforming RF and ultrasonic counterparts at similar distances. The PWM backscatter uplink achieves a data rate of 17.73 kbps with a low bit energy of 0.9 pJ/bit. Furthermore, the system achieves a lower BER of $8.5\times 10^{-5}$ at an extended distance of 5 cm, thanks to the MLP decoding algorithm.

\begin{table*}[]
\caption{\textbf{Comparison with state-of-the-art mm-scale wireless implants with bidirectional telemetry.}}
\label{table}
\centering
\setlength{\tabcolsep}{3.6pt}
\renewcommand{\arraystretch}{1.2}
\begin{tabular}{|cc|c|c|c|c|c|c|c|}
\hline
\multicolumn{2}{|c|}{\textbf{}} &
  \textbf{This work} &
  \begin{tabular}[c]{@{}c@{}}Z. Yu \\ MobiCom'22\cite{yu_magnetoelectric_2022}\end{tabular} &
  \begin{tabular}[c]{@{}c@{}}J. Lim\\ VLSI'21\cite{lim_light-tolerant_2022}\end{tabular} &
  \begin{tabular}[c]{@{}c@{}}J. Lee \\ Nat. Elec.'21\cite{lee_neural_2021}\end{tabular} &
  \begin{tabular}[c]{@{}c@{}}Y. Jia\\ ISSCC'20\cite{jia_trimodal_2020}\end{tabular} &
  \begin{tabular}[c]{@{}c@{}}S. Sonmezoglu\\ ISSCC'20\cite{sonmezoglu_45mm3_2020}\end{tabular} &
  \begin{tabular}[c]{@{}c@{}}M. Ghanbari\\ JSSC'19\cite{ghanbari_sub-mm3_2019}\end{tabular} \\ \hline
\multicolumn{2}{|c|}{\textbf{Biomedical Function}} &
  \textbf{\begin{tabular}[c]{@{}c@{}}Stimulation, \\ Recording\end{tabular}} &
  Stimulation &
  Recording &
  Recording &
  \begin{tabular}[c]{@{}c@{}}Stimulation, \\ Recording\end{tabular} &
  O\textsubscript{2} sensing &
  Recording \\ \hline
\multicolumn{2}{|c|}{\textbf{SoC Technology (nm)}} &
  \textbf{180} &
  180 &
  180 &
  65 &
  350 &
  65 &
  65 \\ \hline
\multicolumn{1}{|c|}{\multirow{2}{*}{\textbf{Power}}} &
  \textbf{Source} &
  \textbf{ME} &
  ME &
  Photodiode &
  RF &
  Inductive &
  Ultrasound &
  Ultrasound \\ \cline{2-9} 
\multicolumn{1}{|c|}{} &
  \textbf{Transfer Efficiency} &
  \textbf{0.37\% (2cm)} &
  0.2\% (2cm) &
  N/A &
  0.08\% (0.8cm) &
  N/A &
  N/A &
  0.06\% (1.8cm) \\ \hline
\multicolumn{1}{|c|}{\multirow{2}{*}{\textbf{\begin{tabular}[c]{@{}c@{}}Downlink \\ Data\end{tabular}}}} &
  \textbf{Modality} &
  \textbf{ME} &
  ME &
  Photodiode &
  RF &
  Inductive &
  \multirow{2}{*}{N/A} &
  \multirow{2}{*}{N/A} \\ \cline{2-7}
\multicolumn{1}{|c|}{} &
  \textbf{Modulation} &
  \textbf{Time} &
  Time &
  PWM &
  PWM &
  OOK &
   &
   \\ \hline
\multicolumn{1}{|c|}{\multirow{7}{*}{\textbf{\begin{tabular}[c]{@{}c@{}}Uplink \\ Data\end{tabular}}}} &
  \textbf{Modality} &
  \textbf{ME} &
  ME &
  LED &
  RF &
  RF &
  Ultrasound &
  Ultrasound \\ \cline{2-9} 
\multicolumn{1}{|c|}{} &
  \textbf{f\textsubscript{carrier} (kHz)} &
  \textbf{331} &
  335 &
  N/A &
  900000 &
  433000 &
  2000 &
  1780 \\ \cline{2-9} 
\multicolumn{1}{|c|}{} &
  \textbf{Modulation} &
  \textbf{\begin{tabular}[c]{@{}c@{}}PWM\\ Backscatter\end{tabular}} &
  \begin{tabular}[c]{@{}c@{}}LSK\\ Backscatter\end{tabular} &
  PGM &
  \begin{tabular}[c]{@{}c@{}}LSK\\ Backscatter\end{tabular} &
  OOK &
  \begin{tabular}[c]{@{}c@{}}LSK\\ Backscatter\end{tabular} &
  \begin{tabular}[c]{@{}c@{}}AM\\ Backscatter\end{tabular} \\ \cline{2-9} 
\multicolumn{1}{|c|}{} &
  \textbf{Data Rate (kbps)} &
  \textbf{17.73} &
  1.6 &
  0.3 &
  10000 &
  6780 &
  60 &
  35 \\ \cline{2-9} 
\multicolumn{1}{|c|}{} &
  \textbf{Data Rate / f\textsubscript{carrier}} &
  \textbf{0.054} &
  0.0047 &
  N/A &
  0.011 &
  0.0156 &
  0.03 &
  0.0197 \\ \cline{2-9} 
\multicolumn{1}{|c|}{} &
  \textbf{Bit Energy (pJ/Bit)} &
  \textbf{0.9} &
  N/A &
  253 &
  N/A &
  1342.2 &
  55 &
  603 \\ \cline{2-9} 
\multicolumn{1}{|c|}{} &
  \textbf{BER} &
  \textbf{$\mathbf{8.5\times10\textsuperscript{-5}}$ (5cm)} &
  $9\times10\textsuperscript{-4}$ (1.5cm) &
  N/A &
  $5\times10\textsuperscript{-3}$ (0.8cm) &
  N/A &
  $9\times10\textsuperscript{-5}$ (5cm) &
  N/A \\ \hline
\multicolumn{2}{|c|}{\textbf{Max Distance (cm)}} &
  \textbf{5} &
  2 &
  N/A &
  0.8 &
  N/A &
  5 &
  5 \\ \hline
\end{tabular}
\end{table*}

\section{Conclusion}

In conclusion, this paper presents a bioelectronic implant platform consisting of a 6.7 mm$^3$ implant and a customized TRX for wireless biosensing and stimulation. Power, downlink, and uplink data are transmitted through ME effects. A novel pulse-width modulated ME backscatter uplink explores switched-capacitor energy extraction to achieve a fast amplitude reduction of over 50\% within 2 ME cycles. A high-order ME model is also developed to better understand the energy extraction interface.
System functionality and performance are validated via \textit{ex vivo} and \textit{in vitro} testing, demonstrating efficient and low-BER uplink communication with up to 5 cm distance and a sufficient data rate for streaming LFP recording.






\ifCLASSOPTIONcaptionsoff
  \newpage
\fi

\bibliography{bib/Bio-Electronics,bib/bstcontrol}
\bibliographystyle{IEEEtran}

\vskip -1\baselineskip
\begin{IEEEbiography}[{\includegraphics[width=1in,height=1.25in,clip,keepaspectratio]{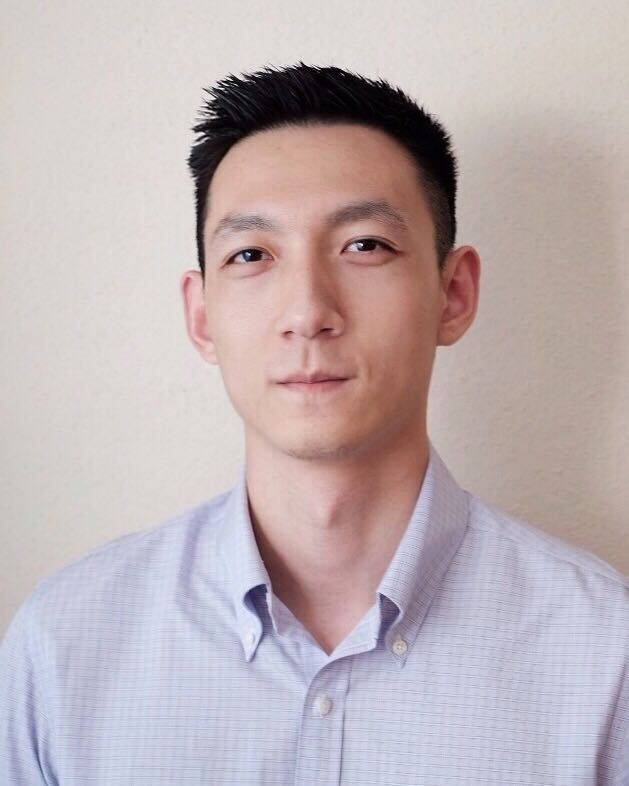}}]{Zhanghao Yu} received a B.E. degree in Integrated Circuit Design and Integrated Systems from the University of Electronic Science and Technology of China, Chengdu, China, in 2016, and an M.S. degree in Electrical Engineering from the University of Southern California, Los Angeles, CA, USA, in 2018. He earned his Ph.D. degree in Electrical and Computer Engineering from Rice University, Houston, TX, USA, in 2023. He worked as an analog design engineering intern at Analog Devices Inc. in 2022 and has been with Intel as an Analog Design Engineer since 2023. His research interests include analog and mixed-signal circuit design for wireless implantable bioelectronics, clocking, power management, wireless power transfer, and low-power communication. Dr. Yu was a recipient of the 2024 SSCS Rising Stars Award and 2021-2022 SSCS Predoctoral Achievement Award. He received Best Paper Awards at 2021 CICC and 2022 MobiCom, and the Best Student Paper Finalist at 2022 RFIC.
\end{IEEEbiography}

\vskip -1\baselineskip
\begin{IEEEbiography}[{\includegraphics[width=1in,height=1.25in,clip,keepaspectratio]{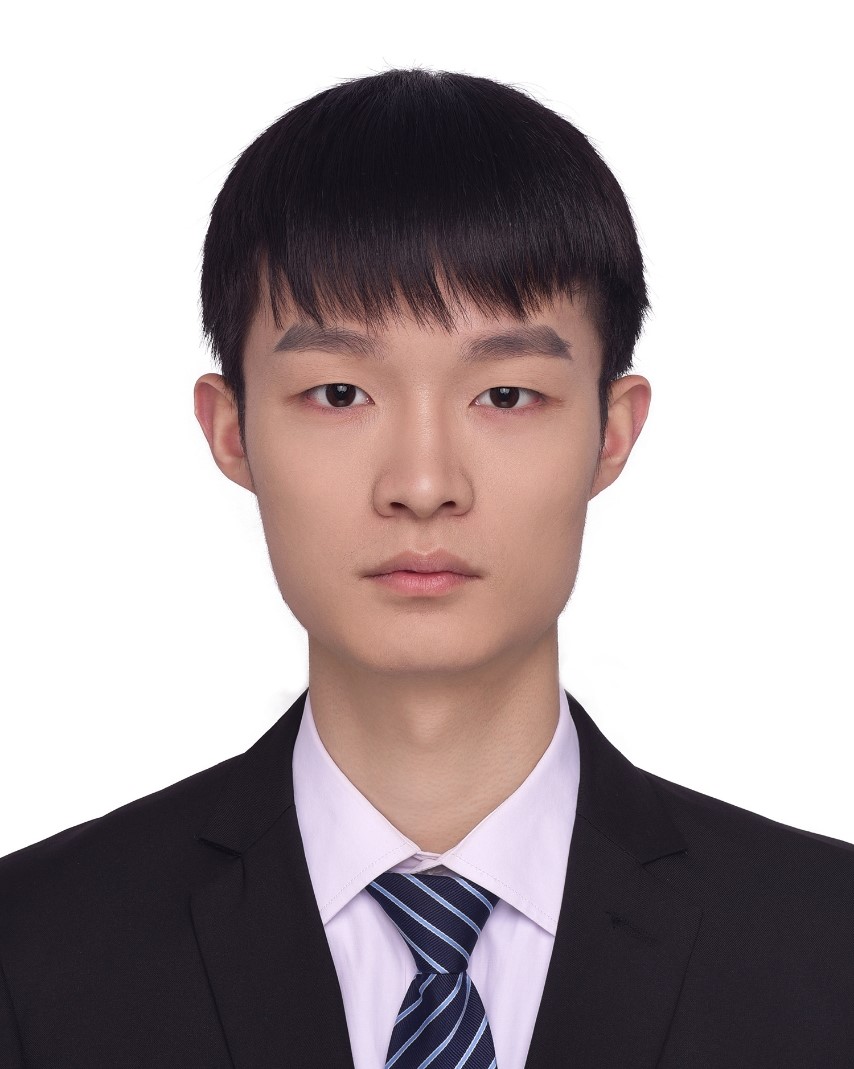}}]{Yiwei Zou} (Graduate Student Member, IEEE) received the B.E. degree in Integrated Circuits and Systems from Huazhong University of Science and Technology, Wuhan, China, in 2022. He is currently working toward his Ph.D. degree in Electrical and Computer Engineering at Rice University, Houston, TX. His research interests include analog and mixed-signal integrated circuits design for power management and bio-electronics.
\end{IEEEbiography}

\vskip -1\baselineskip
\begin{IEEEbiography}[{\includegraphics[width=1in,height=1.25in,clip,keepaspectratio]{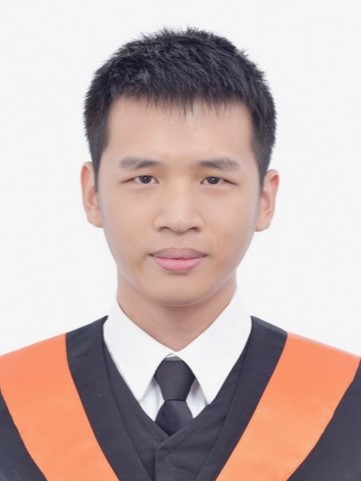}}]{Huan-Cheng Liao} (Graduate Student Member, IEEE)  received the B.S. degree from National Taiwan University, Taipei, Taiwan, in 2020. He is currently working toward the Ph.D. degree in electrical and computer engineering with Rice University, Houston, TX, USA. His current research interests include analog and mixed-signal integrated circuits design.
\end{IEEEbiography}

\vskip -1\baselineskip
\begin{IEEEbiography}[{\includegraphics[width=1in,height=1.25in,clip,keepaspectratio]{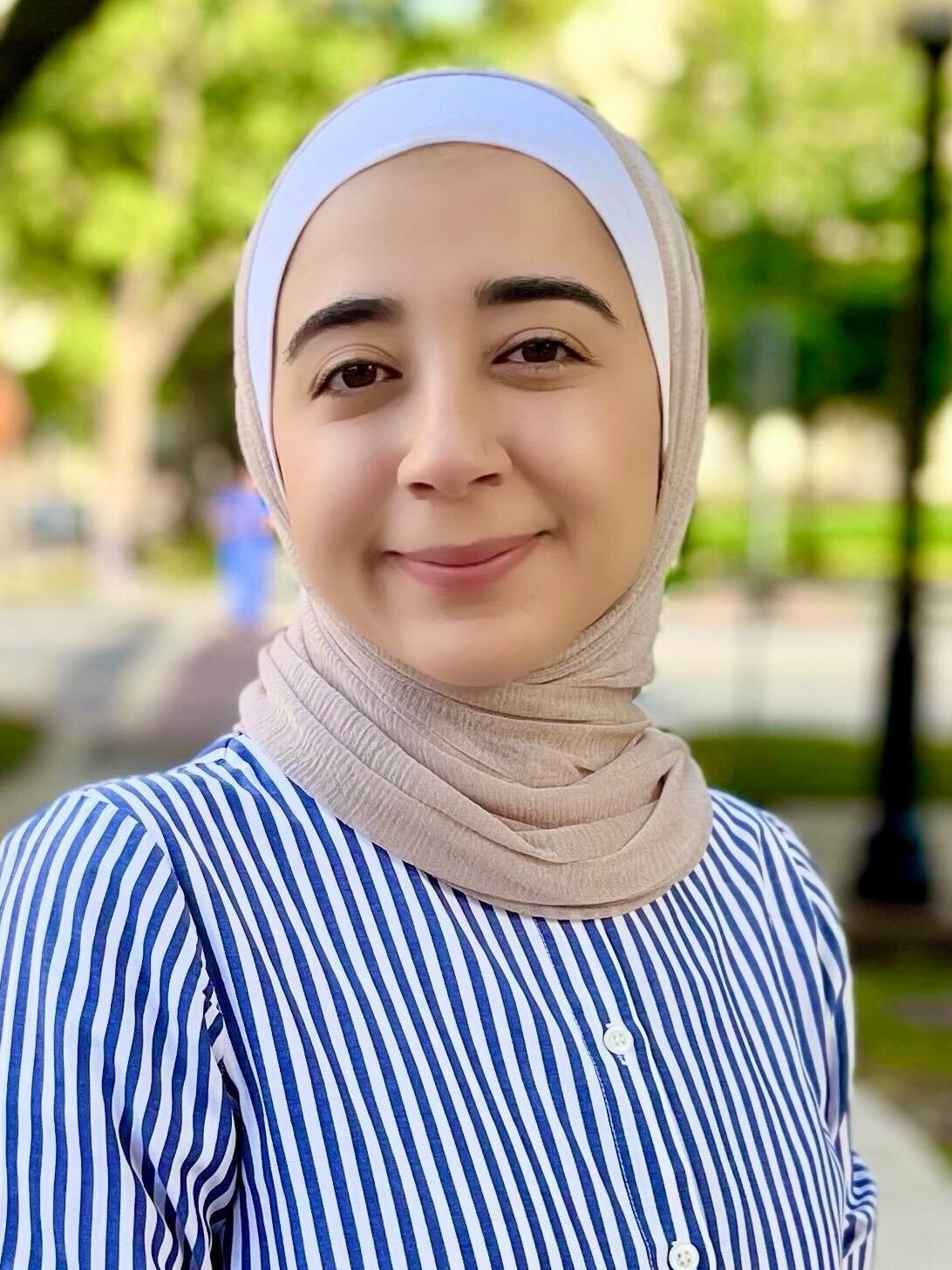}}]{Fatima Alrashdan} is a postdoctoral researcher in the Department of Neuroengineering at Rice University. She earned both her Bachelor's and Master's degrees in Electrical Engineering from Jordan University of Science and Technology in Irbid, Jordan, followed by a PhD in Electrical and Computer Engineering from Rice University in TX, United States. Fatima’s research focuses on developing wireless bioelectronic implants for neuromodulation.
\end{IEEEbiography}

\vskip -1\baselineskip
\begin{IEEEbiography}[{\includegraphics[width=1in,height=1.25in,clip,keepaspectratio]{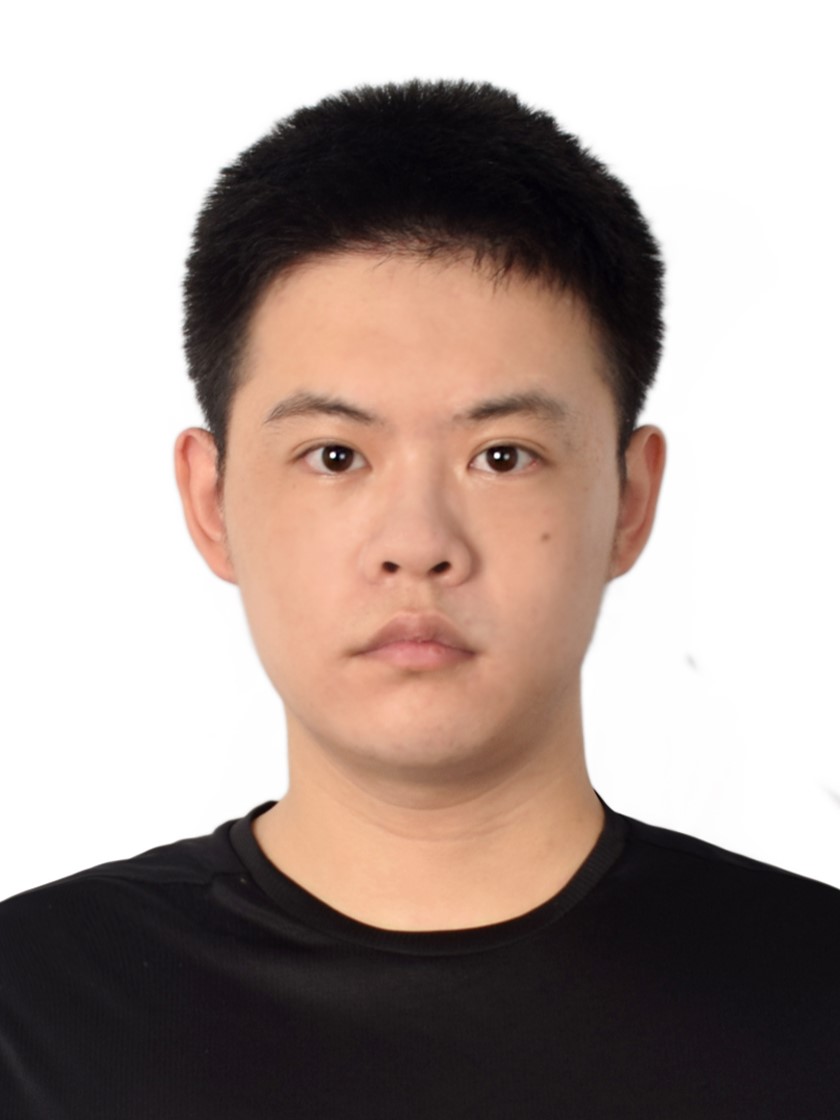}}]{Ziyuan Wen} (Graduate Student Member, IEEE) received a bachelor’s degree in optical and electronic information from the Huazhong University of Science and Technology, Wuhan, China, in 2022. He is currently pursuing a Ph.D. degree in electrical and
computer engineering at Rice University, Houston, TX, USA. His research interests include low-power integrated circuits for streaming data processors, biomedical applications, and in-memory computing accelerators for deep learning.
\end{IEEEbiography}

\vskip -1\baselineskip
\begin{IEEEbiography}[{\includegraphics[width=1in,height=1.25in,clip,keepaspectratio]{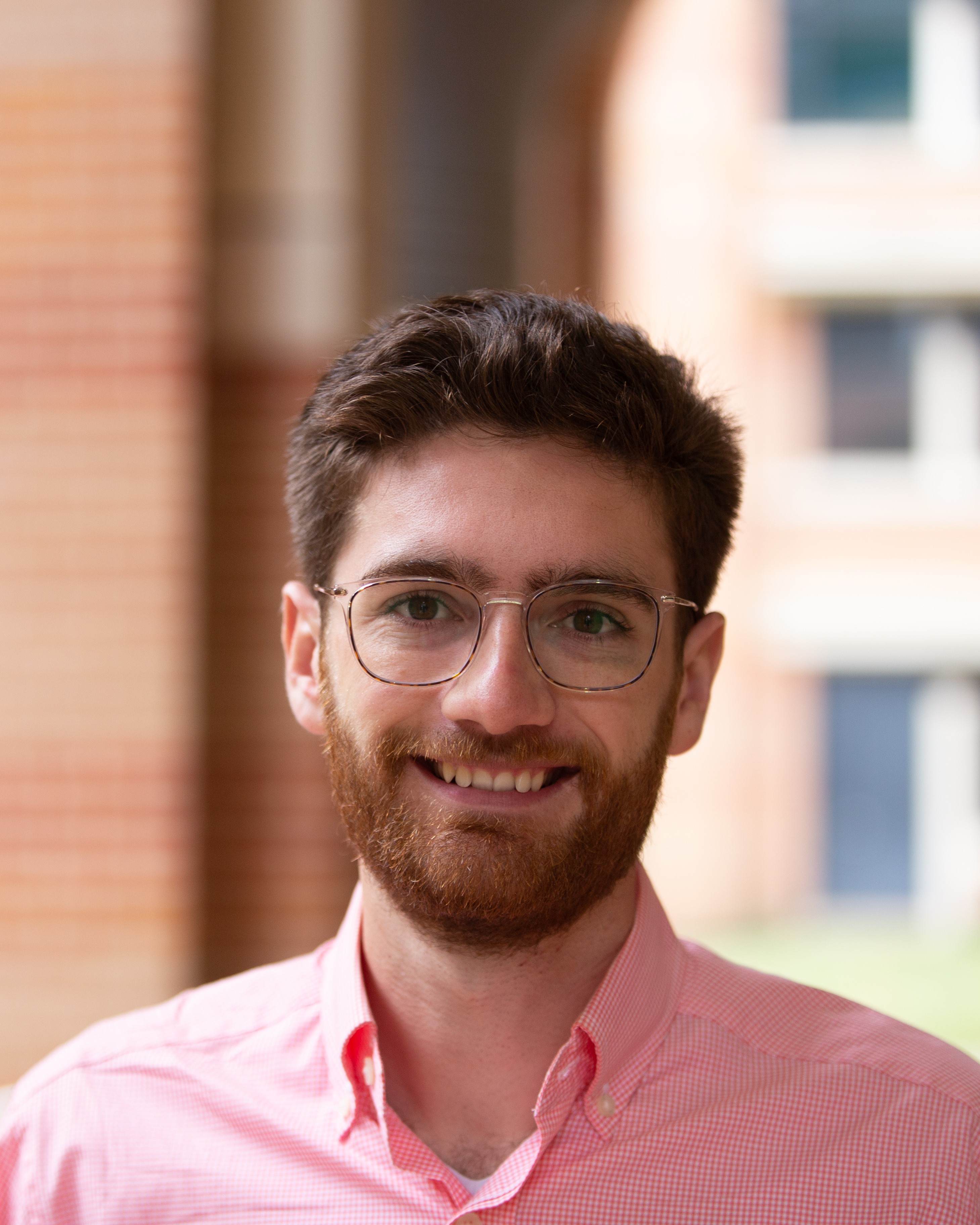}}]{Joshua E. Woods} received a B.S. in Electrical Engineering from the University of Michigan Ann Arbor in 2021. He is currently pursuing a PhD in Electrical Engineering at Rice University in Houston, TX. His research interests include wireless power, embedded systems, and closed-loop bioelectronic systems.
\end{IEEEbiography}

\vskip -1\baselineskip
\begin{IEEEbiography}[{\includegraphics[width=1in,height=1.25in,clip,keepaspectratio]{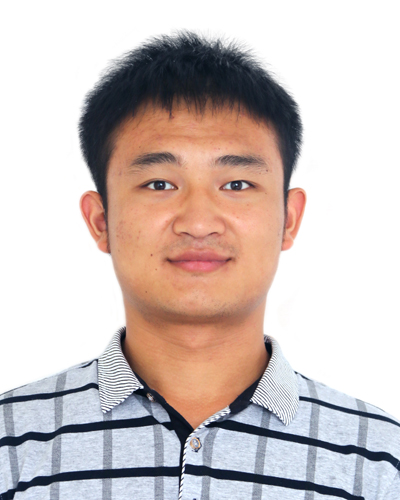}}]{Wei Wang}
(Graduate Student Member, IEEE) received a B.S. degree in electronic information science and technology from the Harbin Institute of Technology, Harbin, China, in 2016, and an M.S. degree in integrated circuit engineering from Tsinghua University, Beijing, China, in 2019. He is currently pursuing the Ph.D. degree in electrical
and computer engineering with Rice University, Houston, TX, USA.
His research interests include mixed-signal circuits and systems design.
\end{IEEEbiography}

\vskip -1\baselineskip
\begin{IEEEbiography}[{\includegraphics[width=1in,height=1.25in,clip,keepaspectratio]{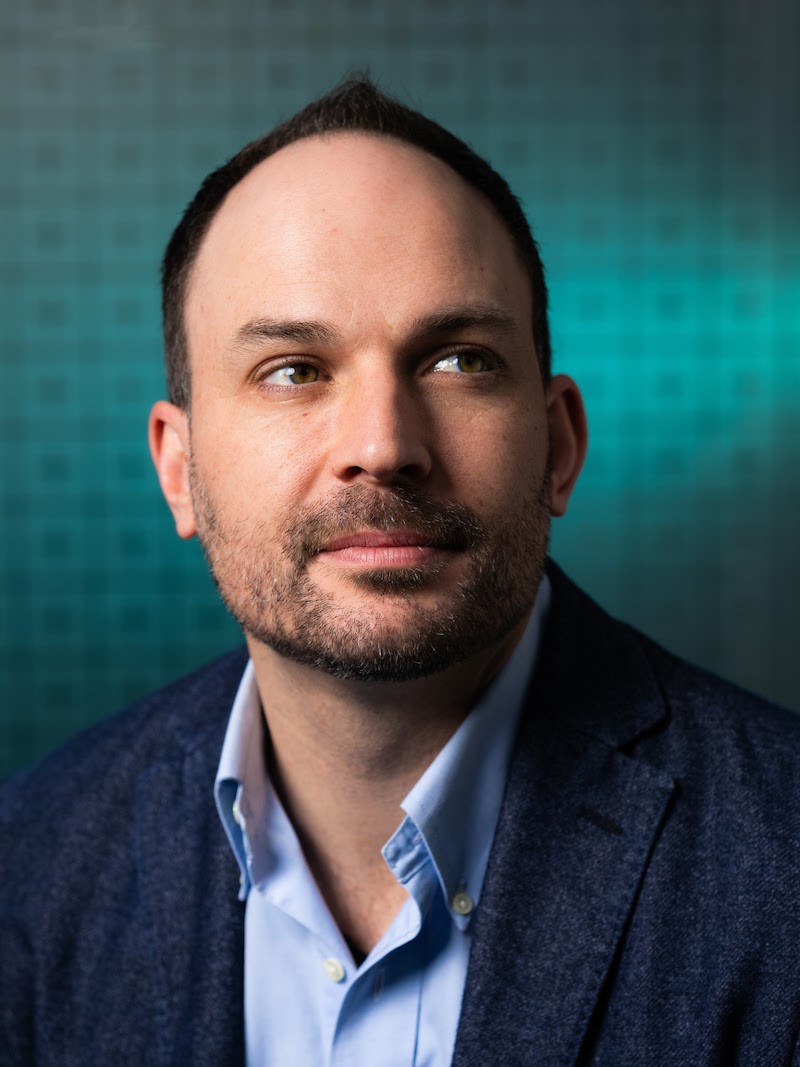}}]
{Jacob Robinson} (Senior Member, IEEE) is a Professor in Electrical \& Computer Engineering and Bioengineering at Rice University where his group develops miniature technologies to manipulate and monitor physiology and neural circuit activity. Prof. Robinson received a B.S. in Physics from UCLA, a Ph.D. in Applied Physics from Cornell University, and completed Postdoctoral training in the Chemistry Department at Harvard. He previously served as the co-chair of the IEEE Brain Initiative and a core member of the IEEE Brain Neuroethics working group. He is currently a member of the IEEE EMBS AdCom and a founding member of the Rice Biotech Launchpad. In addition to his academic work, Dr. Robinson is the co-founder and CEO of Motif Neurotech, which is developing a therapeutic brain computer interface to treat mental health disorders. Dr. Robinson is the recipient of a Charles Duncan Award for Outstanding Academic Achievement, a DARPA Young Faculty Award, and a Materials Today Rising Star Award.
\end{IEEEbiography}

\vskip -1\baselineskip
\begin{IEEEbiography}
[{\includegraphics[width=1in,height=1.25in,clip,keepaspectratio]{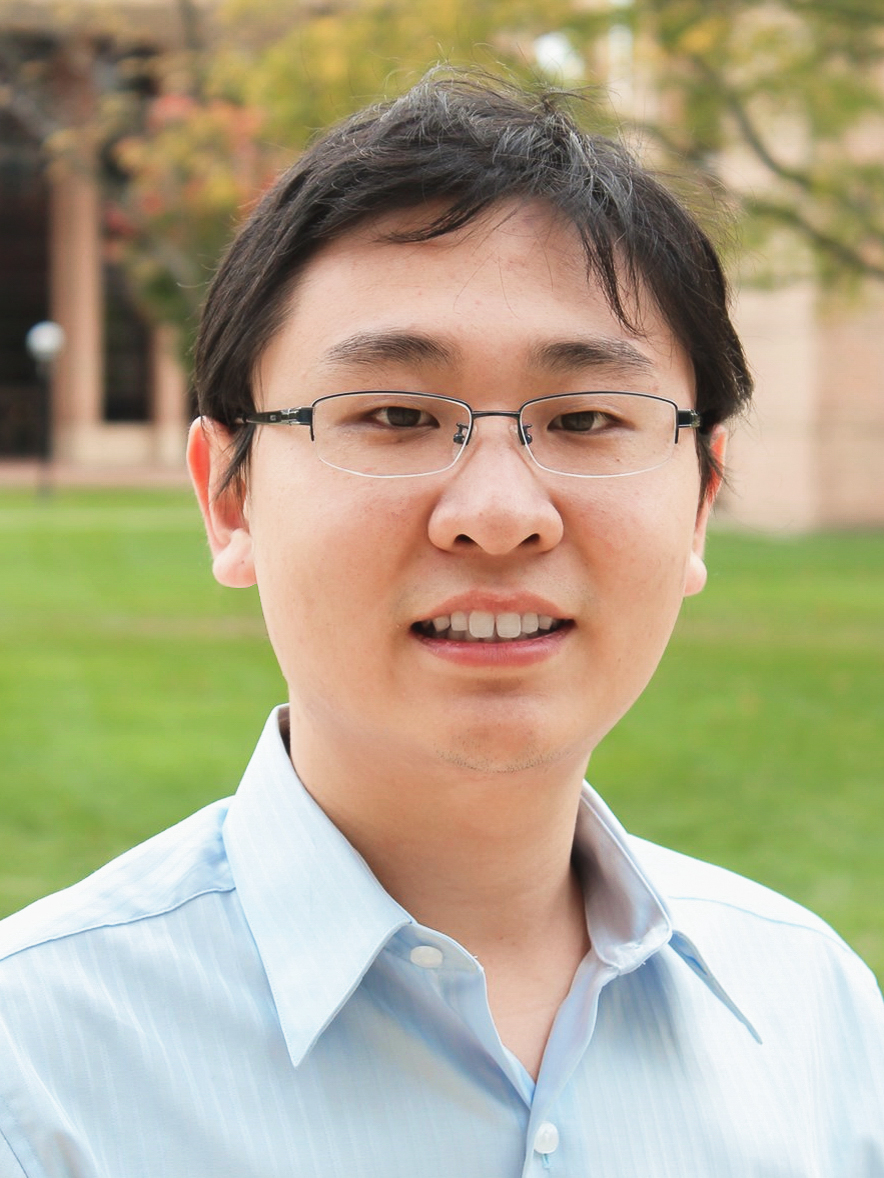}}] {Kaiyuan Yang} (Member, IEEE) is an Associate Professor of Electrical and Computer Engineering at Rice University, USA, where he leads the Secure and Intelligent Micro-Systems (SIMS) lab. He received a B.S. degree in Electronic Engineering from Tsinghua University, China, in 2012, and a Ph.D. degree in Electrical Engineering from the University of Michigan - Ann Arbor, MI, in 2017. His research focuses on low-power integrated circuits and system design for bioelectronics, hardware security, and mixed-signal/in-memory computing. 

Dr. Yang is a recipient of National Science Foundation CAREER Award, IEEE SSCS Predoctoral Achievement Award, and best paper awards from premier conferences in multiple fields, including 2024 Annual International Conference of the IEEE Engineering in Medicine and Biology Society (EMBC), 2022 ACM Annual International Conference on Mobile Computing and Networking (MobiCom), 2021 IEEE Custom Integrated Circuit Conference (CICC), 2016 IEEE International Symposium on Security and Privacy (Oakland), and 2015 IEEE International Symposium on Circuits and Systems (ISCAS). His research was also selected as the research highlights of Communications of ACM and ACM GetMobile magazines, and IEEE Top Picks in Hardware and Embedded Security. He is currently serving as an associate editor of IEEE Transactions on VLSI Systems (TVLSI) and a program committee member of ISSCC, CICC, and DAC conferences. 
\end{IEEEbiography}

\vspace{11pt}

\vspace{11pt}

\vfill

\end{document}